\documentclass[aps,prl,twocolumn,floatfix,groupedaddress,notitlepage,showpacs]{revtex4-1}
\usepackage{graphicx,graphics,epstopdf,color,times,bm,bbm,dsfont,amssymb,amsmath,amsfonts,hyperref,natbib}

\newcommand{\ket}[1]{|#1\rangle}
\newcommand{\Ket}[1]{|#1\rangle\hskip-0.5mm\rangle}
\newcommand{\bra}[1]{\langle#1|}
\newcommand{\Bra}[1]{\langle\hskip-0.5mm\langle#1|}

\newcommand{\BraKet}[2]{\langle\hskip-0.5mm\langle#1|#2\rangle\hskip-0.5mm\rangle}
\newcommand{\ketbra}[2]{|#1\rangle\langle #2|}
\newcommand{\KetBra}[2]{|#1\rangle\hskip-0.5mm\rangle\langle\hskip-0.5mm\langle #2|}

\newcommand{\ignore}[1]{}
\newcommand{\mrm}[1]{\mathrm{#1}}

\let\oldsqrt\sqrt
\def\sqrt{\mathpalette\DHLhksqrt}
\def\DHLhksqrt#1#2{%
\setbox0=\hbox{$#1\oldsqrt{#2\,}$}\dimen0=\ht0
\advance\dimen0-0.2\ht0
\setbox2=\hbox{\vrule height\ht0 depth -\dimen0}%
{\box0\lower0.4pt\box2}}

\begin{document}
\title{Quantum Metrology in Open Systems: Dissipative Cram\'{e}r-Rao Bound}
\author{S. Alipour}
\affiliation{Department of Physics, Sharif University of Technology, Tehran 14588, Iran}
\author{M. Mehboudi}
\affiliation{Department of Physics, Sharif University of Technology, Tehran 14588, Iran}
\author{A. T. Rezakhani}
\affiliation{Department of Physics, Sharif University of Technology, Tehran 14588, Iran}


\begin{abstract}
Estimation of parameters is a pivotal task throughout science and technology. Quantum Cram\'{e}r-Rao bound provides a fundamental limit of precision allowed to achieve under quantum theory. For closed quantum systems, it has been shown how the estimation precision depends on the underlying dynamics. Here, we propose a general formulation for metrology scenarios in open quantum systems, aiming to relate the precision more directly to properties of the underlying dynamics. This feature may be employed to enhance an estimation precision, e.g., by quantum control techniques. Specifically, we derive a Cram\'{e}r-Rao bound for a fairly large class of open system dynamics, which is governed by a (time-dependent) dynamical semi-group map. We illustrate the utility of this scenario through three examples.
\end{abstract}

\maketitle

\textit{Introduction.}---Metrology and parameter estimation lie in the heart of science, and are prevalent in any aspect of technology. The basic task of identification or estimation of a set of unknown parameters essentially requires an inference from a pool of observed data about the parameters or the system to which they are attributed. As errors and imperfections are unavoidable in practice, increasing accuracy of the underlying tasks of data acquisition and inference---hence improving the quality of estimation---is an important goal of metrology \cite{Estimation:book}. Improving quality of measurement instruments and removing potential sources of systematic errors aside, statistics provides useful suggestions for enhancing metrology, such as increasing data size and repeated measurements on an ensemble of $N$ `probe' systems. Additionally (and more interestingly), the underlying physics of the system of interest may also dictate some restrictions or bounds on the ultimate achievable accuracy (usually described through a `Cram\'{e}r-Rao inequality' \cite{Cramer:book}), or even may offer new possibilities to exploit.

In quantum mechanics, measurements act differently than in classical systems. In addition, interactions with an environment or other systems as well as (quantum) correlations can each affect observed data \cite{Helstrom:book}, hence introduce new playing factors in estimation theory. For example, it has been shown that entanglement in a probe ensemble can be exploited to the advantage of a quantum metrology task \cite{Lloyd-qmetrology:PRL}, so that it enables the estimation error of $O(1/N)$ (the ``Heisenberg limit"), in contrast to the classical statistical limit of $O(1/\sqrt{N})$ (the ``shot-noise limit"). Alternatively, enabling $k$-body ($k\geqslant2$) interactions among quantum probe systems has been shown to allow an error of $O(1/\sqrt{N^k})$ \cite{Boixo-etal:PRL07}; or, it has been argued that application of a suitable entangling operator may even offer an error as small as $O(2^{-N})$ \cite{Roy-Braunstein:PRL08} (beyond the Heisenberg limit). Moreover, nonclassicality has been examined as a potential resource for increasing the metrology resolution in quantum optics \cite{Rivas-Luis:PRL10} (for a general framework of resource analysis, see, e.g., Ref.~\cite{Kok:PRL10}). It thus seems natural to expect that some properties of quantum systems can be employed as a useful ``resource" for quantum metrology. Numerous experiments have indeed demonstrated achievability of sub-shot-noise limit error by using aspects of quantum mechanics; see, e.g., Refs.~\cite{experiments}.

In open quantum systems, due to interaction with an environment, the underlying dynamics becomes `noisy.' As a result, formulation and analysis of quantum estimation also becomes more involved \cite{Escher:NatPhys,Watanabe:PRL10}. In general, dynamics of an open system can be described as $\varrho_S(t)=\mathrm{Tr}_{E}[U_{SE}(t,t_0)\varrho_{SE}(t_0)U^{\dag}_{SE}(t,t_0)]$, where $\varrho_{SE}$ is the state of the systems and environment ($SE$), and $U_{SE}(t,t_0)$ is the corresponding unitary evolution \cite{Alicki-Lendi:book,Rivas-Huelga:book}. Thereby one can argue that in general there may exist a flow of information between the system and the environment \cite{Lu-Wang-Sun:PRA10}. Under some conditions, this dynamics can feature quantum Markovian or non-Markovian properties \cite{Alipour:PRA12,Fleming-Hu}. The former case typically appears when the environment has a small decoherence time during which correlations disappear, whereas in the latter correlations (both classical and/or quantum \cite{Pernice:JPB}) with the environment would form and persist. Such correlations are in practice inevitable, which necessitates investigation of noisy quantum metrology \cite{Escher:NatPhys,Lloyd:NP,Guta:NatureC,Adesso:PRA,Monras:PRA,Chin:PRL2012}, and may in turn offer new resources for enhancing estimation tasks. However, developing relatively general frameworks for open-system metrology is still needed and is of fundamental and practical importance.

Here, we first lay out a fairly general formalism for open quantum system metrology. This (re)formulation of the problem (e.g., cf. Ref.~\cite{Escher:NatPhys}) has this advantage that here precision of estimation is more directly related to the underlying dynamics; besides, it is in some sense analogous to the closed system formulation. This formulation also obviates the need for optimization, whereas it provides efficient and reliable estimation of the error scaling with system size, which is always achievable (and often close to exact ultimate precision). Specifically, we derive a quantum Cram\'{e}r-Rao bound (QCRB) for open system dynamics generated through dynamical map with semigroup property. We next illustrate this setting through several examples. The first example shows that how induced correlations of probe quantum systems through a common environment may offer relatively higher precision for estimation in a sense akin to what manybody interactions enable. The other examples provide a comparison of our predicted precision with exact results in two estimation scenarios, which also show some improvement relative to some earlier results.

\textit{Open system dynamics.}---Under some specific conditions, the dynamical equation describing the state of an open system $\varrho_S$ [defined on a Hilbert space $\mathcal{H}_S$] reduces to $\partial_{\tau}\varrho_S(\tau)=\mathcal{L}_{\tau}[\varrho_S(\tau)]$, or equivalently $\varrho_S(\tau)=\mathbf{T}e^{\int_{\tau_0}^{\tau}\mathcal{L}_{\tau'}\mathrm{d}\tau'}[\varrho_S(\tau_0)]$, in which $\mathbf{T}$ denotes time-ordering, and $\mathcal{L}_{\tau}[\circ]=-i[H_S(\tau),\circ]+\sum_{k}\eta_k(\tau) (A_k (\tau) \circ A_k^{\dag}(\tau) -(1/2)\{A_k^{\dag}(\tau) A_k(\tau),\circ\})$ (for some set of operators $\{A_k(\tau)\}$) is the (Lindbladian) generator of the dynamical map, with $H_S(\tau)$ being the system Hamiltonian up to a Lamb shift term (we omit subscript $S$ henceforth). We have also assumed $\hbar\equiv 1$. In (time-dependent) Markovian evolutions, we have $\eta_k (\tau)\geqslant0~\forall k,\tau$; while if some $\eta_k$ becomes negative for some intervals, the associated dynamics would be non-Markovian \cite{Alicki-Lendi:book,Rivas-Huelga:book,Alipour:PRA12,Fleming-Hu}.

Let us assume that a set of unknown parameters $\mathbf{x}=(\mathrm{x}_1,\ldots,\mathrm{x}_{\ell})$ are to be estimated in a quantum system subject to interaction with an environment. For simplicity of our analysis, here we consider the single-parameter case, while generalization of our framework to the multi-parameter case is also straightforward (see example II in the sequel). In the closed-system scenario, this parameter $\mathrm{x}$ is usually assumed to enter into the dynamics as a linear coupling in the Hamiltonian $H(\mathrm{x})=\mathrm{x}\mathsf{H}$ acting on some known initial state. In the open-system scenario, similarly the devised dynamics would in general depend on $\mathrm{x}$ as $\partial_{\tau}{\varrho}(\mathrm{x},t)=\mathcal{L}_{\tau}(\mathrm{x})[\varrho(\mathrm{x},\tau)]$. For our later use, we vectorize this equation, which yields $\partial_{\tau}\Ket{{\varrho}}=\mathcal{\widetilde{L}}_{\tau}(\mathrm{x})\Ket{\varrho}$, where $\widetilde{\mathcal{L}}_{\tau}$ is the matrix representation of $\mathcal{L}_{\tau}$ \cite{Caves-Andersson,vec,SM}. Next we define the normalized pure state $\widetilde{\varrho}\equiv \Ket{\varrho}\Bra{\varrho}/\mathrm{Tr}[\varrho^2]$ (in $\mathcal{H}^{\otimes 2}$), and assume $\widetilde{\mathcal{L}}_\tau(\mathrm{x},\tau) = \mathrm{x(\tau)}\widetilde{\mathsf{L}} $, where $\widetilde{\mathsf{L}}$ does not depend on time; hence
\begin{equation}
\widetilde{\varrho}(\mathrm{x},\tau)=\frac{ e^{\int_0^\tau{\mathrm{x(s)}}\mathrm{d}s~ \widetilde{\mathsf{L}}} ~\widetilde{\varrho}(0)~ e^{\int_0^\tau{\mathrm{x(s)}}\mathrm{d}s~	\widetilde{\mathsf{L}}^ \dagger}}{ \mathrm{Tr}[e^{\int_0^\tau{\mathrm{x(s)}}\mathrm{d}s~ \widetilde{\mathsf{L}}} ~\widetilde{\varrho}(0)~ e^{\int_0^\tau{\mathrm{x(s)}}\mathrm{d}s~ \widetilde{\mathsf{L}}^ \dagger}]}.
\label{dyn-eq}
\end{equation}
The initial preparation $\widetilde{\varrho}(0)$ may itself depend on $\mathrm{x}$, but here we do not assume such generality.

\textit{QCRB for open system metrology.}---Given a data set $\mathcal{D}\equiv\{\gamma_{i}\}$ constituted from some measurement outcomes $\gamma_{i}$ over $N$ (identical) probe systems, an estimator $\mathrm{x}_{\mathrm{est}}(\mathcal{D})$ is chosen for the true value $\mathrm{x}$. By repeating this scenario $M$ times and averaging, the precision of the estimated $\mathrm{x}$, evaluated by $\delta \mathrm{x}=\sqrt{\mathrm{var}(\mathrm{x})}$, is then fundamentally limited by the QCRB
\begin{equation}
\delta\mathrm{x} \geqslant 1/\sqrt{M\mathcal{F}^{(\mathrm{Q})}(\mathrm{x};N)}.
\end{equation}
Here, $\mathrm{var}(\mathrm{x})$ is the variance of any unbiased estimator $\mathrm{x}_{\mathrm{est}}(\mathcal{D})$ (for which, by definition, $\langle \mathrm{x}_{\mathrm{est}}\rangle=\mathrm{x}$, with $\langle \circ\rangle$ denoting the average with respect to the underlying quantum probability distribution), and $\mathcal{F}^{(\mathrm{Q})}(\mathrm{x};N)$ is the so-called ``quantum Fisher information" (QFI) \cite{Braunstein-Caves:QFI,Lloyd:NP,Hayashi:book}. By assuming the state of each $N$-probe set to be $\varrho^{(N)}(\mathrm{x},\tau)$ (hereafter we omit superscript $N$ for brevity) and assigning the corresponding symmetric logarithmic derivative $L_{\varrho}$ through $\partial_{\mathrm{x}}\varrho=(L_{\varrho} \varrho + \varrho L_{\varrho})/2$, the QFI is defined as $\mathcal{F}^{(\mathrm{Q})}(\mathrm{x},\tau;N) \equiv \mathrm{Tr}[\varrho(\mathrm{x},\tau)L^2_{\varrho(\mathrm{x},\tau)}]$.

We remind that in \textit{closed} systems, with $\varrho(\mathrm{x},\tau) =U(\mathrm{x},\tau)\varrho(0)U^{\dag}(\mathrm{x},\tau)$, the spectral decomposition $\varrho=\sum r_i|r_i\rangle \langle r_i|$ and $L_{\varrho}=2 \sum_{ij}\langle r_i|\partial_{\mathrm{x}}\varrho|r_j\rangle/(r_i+r_j) |r_i\rangle \langle r_j|$ (valid for general dynamics) lead to a direct relation between $\mathcal{F}^{(\mathrm{Q})}$ and the interaction $H$. In particular, when $H(\mathrm{x}) \equiv \mathrm{x} \mathsf{H}$ and $\varrho$ is pure, we have
\begin{equation}
\mathcal{F}^{(\mathrm{Q})}=4\tau^2 \mathrm{Cov}_{\varrho}(\mathsf{H},\mathsf{H})
\label{Fisher-H}
\end{equation}
(with equality replaced with $\leqslant$ for mixed $\varrho$), where $\mathrm{Cov}_{\varrho}(X,Y)\equiv \langle XY\rangle_{\varrho} - \langle X\rangle_{\varrho} \langle Y \rangle_{\varrho}$ is the covariance of a pair of observables $X$ and $Y$ with respect to the state $\varrho$, which here is the very quantum standard deviation $\Delta^2 \mathsf{H}$ (with $\langle \circ \rangle_{\varrho}\equiv \mathrm{Tr}[\varrho ~\circ]$). The resulting relation
\begin{equation}
\delta \mathrm{x}\geqslant1/\big(2\tau\sqrt{M}\sqrt{\mathrm{Cov}_{\varrho}(\mathsf{H},\mathsf{H})}\big)= 1/(2\tau\sqrt{M}\Delta \mathsf{H}),
\label{closed}
\end{equation}
where $\Delta \mathsf{H}\equiv \sqrt{\Delta^2 \mathsf{H}}$, is more in the spirit of an uncertainty-like relation \cite{Braunstein-Caves:QFI}, and shows explicitly how the precision is dictated by the interaction. In \textit{open}-system cases, however, deriving similar, direct relation is hardly possible since, e.g., calculating $L_{\varrho(\mathrm{x},\tau)}$
is involved as it requires the knowledge of the spectral decomposition of the density matrix. Thus it is difficult to capture how interaction with an environment affects the QFI and the precision. To partially alleviate this issue, here we follow an alternative approach working with the vectorized state $\widetilde{\varrho}$ instead, which enables a bound somewhat akin to Eq.~(\ref{closed})---with $H$ replaced with $\mathcal{L}$. Although our method gives bounds on the QFI (not its exact value), we demonstrate that this formalism retains significant utility in suggesting correct behavior (e.g., scaling) for the estimation error, and show this explicitly in various examples.

Now from the symmetric logarithmic derivative $L_{\widetilde{\varrho}} = 2\partial_{\mathrm{x}}\widetilde{\varrho}$, one can define an associated QFI $\widetilde{\mathcal{F}}^{(\mathrm{Q})}$ by replacing $(\varrho,L_{\varrho})\to(\widetilde{\varrho},L_{\widetilde{\varrho}})$ in $\mathcal{F}^{(\mathrm{Q})}$. After some straightforward algebra \cite{SM}, using the dynamical equation Eq.~(\ref{dyn-eq}), and assuming a linear $\mathrm{x}$-dependence as $\widetilde{\mathcal{L}}_{\tau}(\mathrm{x})\equiv \mathrm{x}\widetilde{\mathsf{L}}$, it can be seen that
\begin{equation}
\widetilde{\mathcal{F}}^{(\mathrm{Q})} =\frac{4}{\big[\partial_{\tau}\ln\mathrm{x}(\tau)\big]^2}\mathrm{Cov}_{\widetilde{\varrho}} (\widetilde{\mathsf{L}}^{\dagger},\widetilde{\mathsf{L}}),
\label{tildeFisher-Lind}
\end{equation}
which for the time-independent case reduces to $4\tau^2\mathrm{Cov}_{\widetilde{\varrho}} (\widetilde{\mathsf{L}}^{\dagger},\widetilde{\mathsf{L}})$. This relation is analogous to Eq.~(\ref{Fisher-H}), where instead of the Hamiltonian we have the generator of the open dynamics.

The QFI $\widetilde{\mathcal{F}}^{(\mathrm{Q})}$ has a natural interpretation. Recall that $\mathcal{F}^{(\mathrm{Q})}$ indeed emerges from the optimization of the Fisher information over all possible quantum measurements on the system \cite{Braunstein-Caves:QFI}. Similarly then, $\widetilde{\mathcal{F}}^{(\mathrm{Q})}$ is obtained if any quantum measurement on the `system' is allowed. Note, however, that a natural extension of the measurements in $\mathcal{H}$ to $\mathcal{H}^{\otimes 2}$ does not necessarily translate into most general measurements there. For example, a complete set of measurement $\{\Pi_i\}$ (with the properties $\Pi_i\geqslant 0$ and $\sum_i \Pi_i=\openone_{\mathcal{H}}$), when extended simply as $\widetilde{\Pi}_i=\Ket{\Pi_i}\Bra{\Pi_i}$, does not constitute a complete set in the sense that in general $\sum_i \widetilde{\Pi}_i\neq \openone_{\mathcal{H}^{\otimes 2}}$.

Let us see how $\widetilde{\mathcal{F}}^{(\mathrm{Q})}$ compares with $\mathcal{F}^{(\mathrm{Q})}$. First we remark that, from vectorizing the very definition of the symmetric logarithmic derivative, we have $L_{\widetilde{\varrho}}=L_{\varrho}\otimes \openone + \openone \otimes L_{\varrho}^T-\partial_{\mathrm{x}}\ln\mathrm{Tr}[\varrho^2]$. This in turn yields the following expression \cite{SM}:
\begin{eqnarray}
\widetilde{\mathcal{F}}^{(\mathrm{Q})}=\frac{2}{\mathrm{Tr}[\varrho^2]}\Big( \mathrm{Tr}[\varrho L_{\varrho}\varrho L_{\varrho}] + \mathrm{Tr}[\varrho^2 L^2_{\varrho}] -2\frac{(\mathrm{Tr}[\varrho^2 L_{\varrho}])^2}{\mathrm{Tr}[\varrho^2]}\Big).\nonumber\\
\label{FF}
\end{eqnarray}
This form is not yet directly related to $\mathcal{F}^{(\mathrm{Q})}$. However, using $ \lambda_{\min}(X)\mathrm{Tr}[Y]\leqslant\mathrm{Tr}[XY]\leqslant \lambda_{\max}(X)\mathrm{Tr}[Y]$ (valid for any pair of positive matrices $X$ and $Y$) [here $\lambda_{{\min}({\max})}(X)$ denotes the minimum (maximum) eigenvalues of $X$], we obtain
\begin{equation}
\frac{{\mathrm{Tr}[\varrho^2]}}{4\lambda_{\max}(\varrho)} \widetilde{\mathcal{F}}^{(\mathrm{Q})}\leqslant\mathcal{F}^{(\mathrm{Q})}\leqslant\frac{{\mathrm{Tr}[\varrho^2]}}{4\lambda_{\min}(\varrho)} \widetilde{\mathcal{F}}^{(\mathrm{Q})}+F(\varrho),
\label{mid-eq}
\end{equation}
where $F(\varrho)\equiv(\mathrm{Tr}[\varrho^2 L_{\varrho}])^2/\big(\lambda_{\min}(\varrho)\mathrm{Tr}[{\varrho^2}]\big)$. Note that the upper bound would be vacuous when $\lambda_{\min}(\varrho)=0$, thus this case needs special care if one wants to use this bound. Another special case is when the evolution is unitary with a pure initial state, i.e., $|\Psi(\mathrm{x},\tau)\rangle=U(\mathrm{x},\tau)|\Psi(0)\rangle$. Here, however, a significant simplification occurs due to $\langle \Psi(\mathrm{x},\tau)|L_{\Psi}|\Psi(\mathrm{x},\tau)\rangle =0$, whence Eq.~(\ref{FF}) reduces to $\widetilde{\mathcal{F}}^{(\mathrm{Q})} = 2 \mathcal{F}^{(\mathrm{Q})}$ (whereas the lower bound of Eq.~(\ref{mid-eq}) gives $\widetilde{\mathcal{F}}^{(\mathrm{Q})} \leqslant 4\mathcal{F}^{(\mathrm{Q})}$).

Equation (\ref{mid-eq}) provides lower and upper bounds on the exact QFI $\mathcal{F}^{(\mathrm{Q})}$. To obtain the scaling of $\mathcal{F}^{(\mathrm{Q})}$, it suffices to find the scaling of the lower bound of Eq.~(\ref{mid-eq}), since if this bound scales as $O(N^p)$ (for some $p\geqslant0$), it is guaranteed that $\mathcal{F}^{(\mathrm{Q})}=O(N^q)$ with some $q\geqslant p$. However, an upper bound on the QFI might result in an unachievable (hence unreliable) estimation error, thus care must be taken with such bounds. This is another distinctive feature of our method in comparison to the methods of Refs.~\cite{Escher:NatPhys,Guta:NatureC} that here we use a lower bound on the QFI to predict the scaling of the estimation error.

Putting everything together, in general we have obtained
\begin{equation}
1/\mathcal{F}^{(\mathrm{Q})} \leqslant \mathcal{K}/\widetilde{\mathcal{F}}^{(\mathrm{Q})},
\label{F-Ftilde}
\end{equation}
where
\begin{align}
\mathcal{K}(\varrho)\equiv \begin{cases} 4\lambda_{\max}(\varrho)/{\mathrm{Tr}[\varrho^2]},~~~\text{$\varrho$ mixed}\\
2,\hskip25mm\text{$\varrho$ pure} \end{cases}
\end{align}
and for the latter case the inequality in Eq.~(\ref{F-Ftilde}) is replaced with equality. This bound only needs the knowledge of the generators of the dynamics ($\mathcal{L}_{\tau}$) and the instantaneous state ($\varrho$), without need to calculate $L_{\varrho}$ or to do any optimization.

A desirable property of $\mathcal{F}^{(\mathrm{Q})}$ is that for a fully product/separable estimation scenario with $N$ product input states, we have $\mathcal{F}^{(\mathrm{Q})}(\mathrm{x},\tau; N)= N \mathcal{F}^{(\mathrm{Q})}(\mathrm{x},\tau;1)$, which naturally carries over to $\widetilde{\mathcal{F}}^{(\mathrm{Q})}(\mathrm{x},\tau;N)$ \cite{SM}. Thus for this special case, at the left- and right-hand sides of Eq.~(\ref{mid-eq}), we must replace $\varrho$ [the $N$-probe state] with $\varrho^{(1)}$ [the single-probe state] and $\widetilde{\mathcal{F}}^{(\mathrm{Q})}(\mathrm{x},\tau;N)$ with $N\widetilde{\mathcal{F}}^{(\mathrm{Q})}(\mathrm{x},\tau;1)$, which exhibits the expected shot-noise scaling $O(1/\sqrt{N})$ for the estimation error.

\begin{figure}[tp]
\includegraphics[scale=0.3]{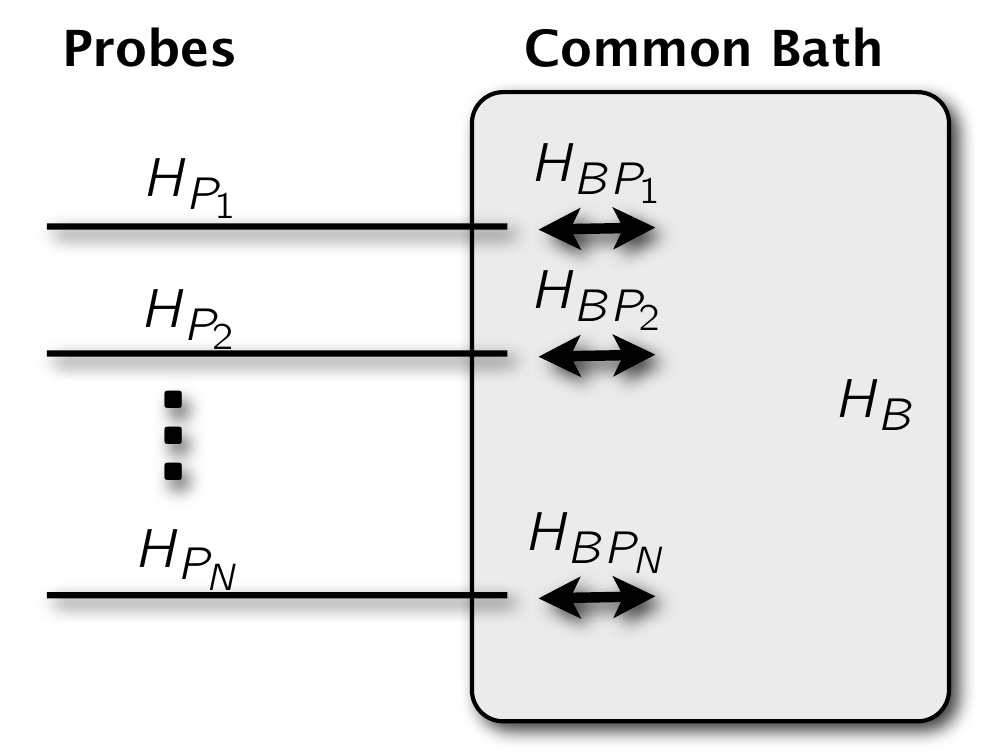}
\caption{$N$ probes, initially well isolated from each other, all interact with a common bath through two-body interactions $H_{BP_i}$. Here $H_{P_i}$ and $H_B$ are the free Hamiltonians of probe $i$ and the bath, respectively. These two-body interactions may induce a manybody quantum correlation among the probes \cite{Braun:PRL02,Benatti:PRL03}.}
\label{fig:kbody}
\end{figure}

\textit{Example I.}---We assume $N$ probe particles each of which only interacts with a common bath such that the interactions induce all possible $k$-body terms (Fig.~\ref{fig:kbody}) \cite{Braun:PRL02} in the Lindbladian as follow:
\begin{equation}
\mathcal{L}_{\tau}[\varrho]=\sum_{i_1\cdots i_k}\sigma_{i_1}\cdots\sigma_{i_k}~\varrho~ \sigma_{i_1}\cdots \sigma_{i_k}-C_{N,k}~\varrho,
\end{equation}
where $\sigma_{i_j}$ are all the same Pauli matrix (e.g., $\sigma^z$), subscript $i_j$ is the particle index, and the factor $C_{N,k}=\binom{N}{k}$ counts the number of $k$-body operators. This is a generalization of the scenario considered in the closed-system context of Ref.~\cite{Boixo-etal:PRL07}, and is beyond the scope of the analysis in Ref.~\cite{Guta:NatureC} for estimation scenarios with separable channels.

We choose the initial state of the whole $N$-probe system to be the maximally entangled pure state $\varrho(0)=\ketbra{\Psi}{\Psi}$, where $\ket{\Psi}=(\ket{E_{M}}^{\otimes N}-\ket{E_{m}}^{\otimes N})/\sqrt{2}$, and $E_{m}$ ($E_M$) is the smallest (largest) eigenvalue of $\sigma$. For odd $k$s, $\sigma_{i_1}\sigma_{i_2}\cdots \sigma_{i_k}\otimes \sigma_{i_1}\sigma_{i_2}\cdots \sigma_{i_k}(\ket{\Psi}\otimes\ket{\Psi^{\ast}})=\ket{\Psi^{\perp}}\ket{\Psi^{{\perp}^{\ast}}}$, where $\ket{\Psi^{\perp}}=(\ket{E_{M}}^{\otimes N}+\ket{E_{m}}^{\otimes N})/\sqrt{2}$. It is straightforward to see that \cite{SM} $(\ket{\Psi^{\perp}}\ket{\Psi^{{\perp}^{\ast}}}-\ket{\Psi}\ket{\Psi^{\ast}})/\sqrt{2}$ is a normalized eigenvector of $\widetilde{\mathsf{L}}$ corresponding to the eigenvalue $-2C_{N,k}$, whence
\begin{equation}
\frac{\mathcal{K}}{\widetilde{\mathcal{F}}^{(\mathrm{Q})}}=\frac{(e^{-2C_{N,k} \tau \mathrm{x}} + 1)(e^{-4C_{N,k} \tau \mathrm{x}} + 1)}{4\tau^2 C_{N,k}^2e^{-4C_{N,k} \tau \mathrm{x}}}.
\end{equation}
An immediate implication of this relation and that $C_{N,k}=O(N^k)$ is that for small values of the $\mathrm{x}$ parameter a polynomial precision in the estimation can be achieved.

\textit{Example II.}---Consider a dephasing channel acting separately on an $N$-qubit system, described by $\mathcal{L}_{\tau}[\varrho]=i\mathrm{x}_1[H,\varrho] + (1/2)\mathrm{x}_2(\tau)\big(\sum_{m=1}^{N} \sigma^z_{m}\varrho \sigma^z_{m}-N\varrho \big)$, in which $\mathrm{x}_1$ is the gap of the Hamiltonian $H=\sum_{m=1}^{N}|1\rangle_{m}\langle 1|$, whose ground-state energy is zero \cite{Huelga:PRL1997}. We assume two different initial states; the product state $\ket{\Psi_{\mathrm{p}}}=[(\ket{0}+\ket{1})/\sqrt{2}]^{\otimes N}$ and the entangled ``GHZ" state $\ket{\Psi_{\mathrm{e}}} = (|0\rangle^{\otimes N} + |1\rangle^{\otimes N})/\sqrt{2}$.

\begin{figure}[tp]
\includegraphics[scale=.45]{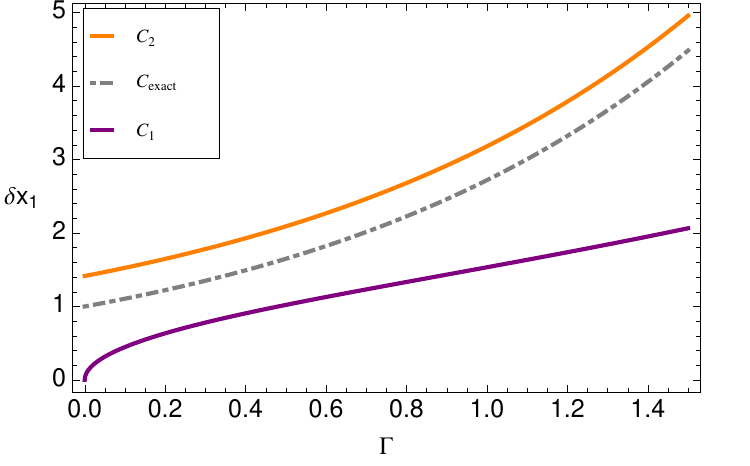}
\caption{(color online.) Factor $C(\Gamma)$ in the scaling $O(C/\sqrt{N})$ of the $\mathrm{x}_1$ estimation with product states. The values $C_1$ [down, purple], $C_{\mathrm{exact}}$ [middle, dot-dashed], and $C_2$ [up, orange] are given through Refs. [10,19], exact calculation, and our method, respectively.}
\label{fig:exm-2-comp}
\end{figure}

\textit{Estimation of $\mathrm{x}_1$.}---Using Eq.~(\ref{F-Ftilde}) and after some algebra \cite{SM}, it is obtained that in the case of product (``$\mathrm{p}$") and entangled (``$\mathrm{e}$") states we have
\begin{eqnarray}
\frac{\widetilde{\mathcal{F}}_{\mathrm{p}}^{(\mrm Q)}(\mathrm{x}_1)}{\mathcal{K} }&=&\frac{N \tau^2 e^{-3\Gamma/2}}{ 2~\mathrm{ch}(\Gamma /2)},\label{pp1}\\
\frac{\widetilde{\mathcal{F}}_{\mathrm{e}}^{(\mrm Q)}(\mathrm{x}_1)}{\mathcal{K}} &=&\frac{N^2\tau^2 e^{-3N\Gamma/2}}{2~\mathrm{ch}(N\Gamma/2)}, \label{ee1}
\end{eqnarray}
whereas the exact QFIs are argued to be \cite{Chin:PRL2012,Campo:PRL2013}
 \begin{eqnarray}
\mathcal{F}_{\mathrm p}^{(\mathrm Q)}(\mathrm{x}_1)&=&N \tau^2e^{-2\Gamma}, \label{p1}\\
\mathcal{F}_{\mathrm e}^{(\mathrm Q)} (\mathrm{x}_1)&=&N^2 \tau^2 e^{-2N\Gamma}.\label{e1}
 \end{eqnarray}
Here $ \Gamma(\tau)\equiv\int_0^{\tau}\mathrm{x}_2(s)\mathrm{d}s$, $\mathrm{ch}=\mathrm{cosh}$, and $\mathrm{sh}=\mathrm{sinh}$. It is evident that the ratio of the bound (\ref{pp1}) and the exact value (\ref{p1}) is always equal to $(1+e^{-\Gamma})^{-1}$; and for large $N$s, the ratio of the bound (\ref{ee1}) and the exact value (\ref{e1}) goes to $1$. Note that when $\mathrm{x}_2=0$, the ratios both are $1/2$, which is consistent with what we expect in the unitary case [$\widetilde{\mathcal{F}}^{(\mathrm{Q})}=2\mathcal{F}^{(\mathrm{Q})}$]. Therefore, our framework correctly captures the scaling of the error in this example. The very problem of estimating $\mathrm{x}_1$ (with product input states) has already been discussed in Refs.~\cite{Escher:NatPhys,Guta:NatureC} too, where it has been found that $\delta\mathrm{x}_1=O(C/\sqrt{N})$, with a given constant $C$. However, interestingly, here our formalism gives a more improved scaling in that it compares with the exact solution more favorably and with a better $C$; see Fig.~\ref{fig:exm-2-comp} and the discussion in Ref.~\cite{SM}.

A more exhaustive comparison of the ``$\mathrm{p}$" and ``$\mathrm{e}$" scenarios necessitates finding optimized measurement times for either. We have performed this analysis in Ref.~\cite{SM} and shown that in the Markovian case of this estimation task no relative advantage is offered by the ``$\mathrm{e}$" scenario, although with a different noise model in the Markovian case the ``$\mathrm{e}$" scenario has been shown to be advantageous \cite{PRL:chaves}. In the non-Markovian case, however, here the ``$\mathrm{e}$" scenario may outperform ``$\mathrm{p}$" for our specific noise model. This sort of comparative analysis, introduced in Ref.~\cite{Chin:PRL2012}, can have intimate implications on experimental realizations.

\textit{Estimation of $\mathrm{x}_2$}.---Similar calculations \cite{SM} yield
\begin{eqnarray}
\frac{\widetilde{\mathcal{F}}_{\mathrm{p}}^{(\mrm Q)}(\mathrm{x}_2)}{\mathcal{K}} &=&
\frac{N e^{-\Gamma/2}}{4[\partial_{\tau}\ln\mathrm{x}_2]^2 ~\mathrm{ch}(\Gamma)\mathrm{ch}(\Gamma/2)},\label{ep1}\\
\frac{\widetilde{\mathcal{F}}_{\mathrm{e}}^{(\mrm Q)}(\mathrm{x}_2)}{\mathcal{K}} &= &\frac{N^2 e^{-N\Gamma/2}}{4[\partial_{\tau}\ln\mathrm{x}_2]^2 ~\mathrm{ch}(N\Gamma)\mathrm{ch}(N\Gamma/2)}.\label{ep2}
\end{eqnarray}
On the other hand, here the exact QFIs are obtained as
\begin{align}
&\mathcal{F}^{\mathrm{(Q)}}_{\mathrm{p}}(\mathrm{x}_2) =\frac{1}{[\partial_{\tau}\ln\mathrm{x}_2]^2}\frac{N e^{-\Gamma}}{2~\mathrm{sh}(\Gamma)},\label{ep11}\\
&\mathcal{F}^{\mathrm{(Q)}}_{\mathrm{e}}(\mathrm{x}_2)=\frac{1}{[\partial_{\tau}\ln\mathrm{x}_2]^2}\frac{N^2 e^{-N\Gamma}}{2~\mathrm{sh}(N\Gamma)}. \label{ep12}
\end{align}
Again it is evident that the ratio of the bound (\ref{ep1}) and the exact value (\ref{ep11}) is always $(e^{2\Gamma}-e^{\Gamma})/(e^{2\Gamma}+1)$; and the ratio of the bound (\ref{ep2}) to the exact value (\ref{ep12}), for large $N$s, goes to $1$. These results also exhibit correct scalings and behaviors.

\textit{Example III}.---Consider a lossy bosonic channel described by $\mathcal{L}_{\tau}[\varrho]= \mrm x[a\varrho a^{\dagger} -( \widehat{n}\varrho + \varrho \widehat{n})/2]$, where $a$ ($a^{\dag}$) is the bosonic annihilation (creation) operator, $\widehat{n}=a^{\dag}a$, and $\mathrm{x}$ is the loss parameter. The QCRB for estimation of $\varphi$---defined through $\tan^2[\varphi(\mrm x,\tau)] = e^{\mrm x \tau} - 1$---has been obtained as $\delta\varphi\geqslant 1/\sqrt{4\overline{n}\tau}$, and whereby $\delta\mathrm{x}\geqslant \sqrt{\mathrm{x}/(\overline{n}\tau)}$, where $\overline{n}=\mathrm{Tr}[\widehat{n}\varrho(0)]$ \cite{Paris:PRL}. Particularly, it has been shown that Fock states are optimal for this estimation \cite{Adesso:PRA}. Here we revisit this example and demonstrate that the behavior of the error is captured correctly in our framework.

The evolution of this system, when the initial state is $\varrho(0)=|N\rangle \langle N|$ (whence $\overline{n}=N$), is given by
\begin{equation}
\Ket{\varrho(\mrm x, \tau)} =\sum_{m=0}^{N} \mathrm{s}^{2m}\mathrm{c}^{2(N-m)}C_{N,m}|N-m\rangle |N-m\rangle
\end{equation}
in which $\mathrm{s}=\sin\varphi$ and $\mathrm{c}=\cos\varphi$.

\begin{figure}[tp]
\includegraphics[scale=0.55]{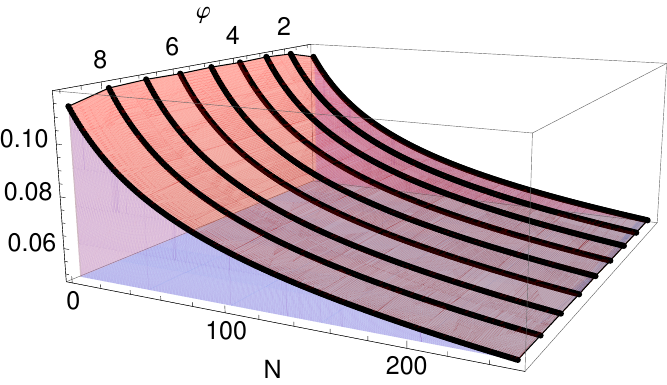}
\caption{$(\delta\varphi)_{\min}$ [or $\mathcal{K}/\widetilde{\mathcal{F}}^{(\mathrm{Q})}$] vs. $\varphi$ and $N$ in example III [Eq.~(\ref{eq:formula})]. Black curves represent values $\varphi\in\{\pi/20,2\pi/20,\ldots,9\pi/20\}$ all showing the scaling $c(\varphi)/\sqrt{N}$, where $c(\varphi)\approx 1/2$ as in Ref.~\cite{Paris:PRL}.}
\label{fig:pl}
\end{figure}
The analytic expression of $\mathcal{K}/\widetilde{\mathcal{F}}^{(\mathrm{Q})}$ can be found as
\begin{eqnarray}
&&\frac{\mathcal{K}}{\widetilde{\mathcal{F}}^{(\mathrm{Q})} (\varphi)} = (1/4)\cot^{2}\varphi ~\underset{0\leqslant m\leqslant N}{\max} \big[C_{N,m}\mathrm{s}^{2m} \mathrm{c}^{2(N-m)}\big]\times\nonumber\\
&&~\Big[\sum _{m=0}^N \mathrm{s}^{4m}\mathrm{c}^{4(N-m)}C_{N,m}^2 A_{N,m}^2\nonumber\\
&&~-\frac{\left(\sum _{m=0}^N \mathrm{s}^{4m}\mathrm{c}^{4(N-m)}C_{N,m}^2 A_{N,m}\right)^2}{\sum_{m=0}^{N} \mathrm{s}^{4m} \mathrm{c}^{4(N-m)} C_{N,m}^2}\Big]^{-1},
\label{eq:formula}
\end{eqnarray}
where $A_{N,m}=m(1+\cot^2\varphi)-N$. Using $\mathcal{F}^{(\mathrm{Q})}(\mrm x) = (\partial_{\mrm x}\varphi)^2 \mathcal{F}^{(\mathrm{Q})}(\varphi)$, one can relate the lower bound for estimation of $\mrm x$ to that of $\varphi$. Figure \ref{fig:pl} depicts $(\delta\varphi)_{\min}$, which verifies that our bound gives the correct behavior of the error.

\textit{Summary and outlook.}---Here we have outlined a fairly general formalism for open quantum system metrology. In this formulation, the precision of estimation is more directly related to the underlying dynamics, in some sense similar to the closed-system formulation. This property may enable to enhance metrology in noisy systems by employing quantum/classical control methods to partially engineer or manipulate the system. We have derived a quantum Cram\'{e}r-Rao bound for open system dynamics generated through dynamical map with the semigroup property. It has been shown that this method always gives an achievable precision (which is mostly close or equal to ultimate bound), while it also offers other advantages, such as providing an efficient method for deriving bounds based on dynamics, over existing methods. This setting was then illustrated through some examples. The first example implied possibility of exploiting induced correlations of probe quantum systems through a common environment in order to achieve a relatively higher precision. Other two examples have illustrated that our bound could indeed give correct scaling of the estimation error.

Our formalism may introduce novel methods for utilizing some of the resources offered in open quantum dynamics, such as induced manybody correlations and memory, to hopefully enhance a quantum estimation task in the presence of noise. This in turn can spur applications in, e.g., quantum sensing \cite{Lloyd:NP,GLM-new:PRL} and quantum control of optomechanical devices for advanced means and technologies \cite{opt}.

\textit{Acknowledgments.---}Supported by Sharif University of Technology's Office of Vice-President for Research.


\appendix
\begin{widetext}
\section{Supplemental Material:}
\section{Vectorization}
\label{app:vec}

Let $\{|i\rangle\}$ be a given orthonormal basis for the Hilbert space $\mathcal{H}$. To an arbitrary linear operator
\begin{equation}
A=\sum_{ij}\langle i|A|j\rangle |i\rangle\langle j|\in\mathcal{S}(\mathcal{H}),
\end{equation}
one can assign a vector
\begin{equation}
\Ket{A}=\sum_{ij}\langle i|A|j\rangle|i\rangle |j\rangle\in \mathcal{S}(\mathcal{H}^{\otimes 2}).
\end{equation}
This (row) ``vectorization" is basis-dependent. To remove any ambiguity, henceforth we use the ``computational basis," in which $|i\rangle\overset{.}{=}(0\ldots 0~~1~~0\ldots 0)^T$. In the examples described in the next sections, though, we may choose the more natural photon-number basis for the vectorization.

\begin{enumerate}

\item For an arbitrary Hermitian operator $A$ with the spectral decomposition $A=\sum_{i}a_i|a_i\rangle \langle a_i|$ (where $a_i\in\mathbb{R}$ and $\langle a_i|a_j \rangle=\delta_{ij}$), we have
\begin{equation}
\Ket{A}=\sum_{i}a_i|a_i\rangle |a^*_i\rangle,
\label{id-spectral}
\end{equation}
where $|a^*_i\rangle$ is the complex conjugate of $|a_i\rangle$ in the computational basis.

\textit{Proof.} Assume $|a_i\rangle=\sum_{i'}\alpha_{ii'}|i'\rangle$, where $\{|i'\rangle\}$ is the computational basis. Thus in the computational basis one can write $A=\sum_{ii'i''}a_i\alpha_{ii'}\alpha_{ii''}^*|i'\rangle \langle i''|$, whence
\begin{eqnarray}
\Ket{A} &=& \sum_{ii'i''}a_i\alpha_{ii'}\alpha^*_{ii''}|i'\rangle |i''\rangle \nonumber\\
&=& \sum_{i}a_i \sum_{i'}\alpha_{ii'}|i'\rangle \sum_{i''}\alpha^*_{ii''} |i''\rangle \nonumber\\
&=& \sum_{i}a_i|a_i\rangle |a^*_i\rangle.\nonumber
\end{eqnarray}
In a similar vein, one can show that for an arbitrary operator $A=\sum_{ij}\langle u_i|A|u_j\rangle |u_i\rangle \langle u_j|$, where $\{|u_i\rangle\}$ is some orthonormal basis (not necessarily the computational basis), the vectorization yields
\begin{equation}
\Ket{A} = \sum_{ij}\langle u_i|A|u_j\rangle |u_i\rangle |u^*_j\rangle,
\end{equation}
where $|u^*_j\rangle$ is the complex conjugate of $|u_i\rangle$ in the computational basis.

\item This Hilbert-Schmidt space $\mathcal{H}_{\text{HS}}$ naturally carries an inner product structure through
\begin{equation}
\BraKet{A}{B}\equiv \mathrm{Tr}[A^{\dag}B].
\label{id-tr}
\end{equation}
Note that although $\Ket{A}$ depends on basis, $\BraKet{A}{B}$ does not.

\textit{Proof.} $\BraKet{A}{B}=\sum_{ij,i' j'}A^*_{ij}B_{i'j'}\langle ij|i'j'\rangle=\sum_{ij}A^*_{ij}B_{ij}=\mathrm{Tr}[A^{\dag}B]$, where $A_{ij}=\langle i|A|j\rangle$.

\item Another vectorization identity that proves useful is as follows:
\begin{equation}
\Ket{ABC}= (A\otimes C^{T})\Ket{B},~\forall A,B,C.
\label{id-tensor}
\end{equation}

\textit{Proof}. This is straightforward by definition,
\begin{eqnarray}
\Ket{ABC}&=&\sum_{ijkl}\langle i |A|j\rangle\langle j |B|k\rangle\langle k|C|l\rangle|i\rangle|l\rangle \nonumber\\
&=& \sum_{ijkl}|il\rangle \langle i l|(A\otimes C^T)|jk\rangle \langle j|B|k\rangle\nonumber\\
&=& \sum_{ij}(A\otimes C^T)|j\rangle |k\rangle \langle j|B|k\rangle\nonumber\\
&=& (A\otimes C^T)\sum_{ij} \langle j|B|k\rangle |j\rangle |k\rangle\nonumber\\
&=&  (A\otimes C^T)\Ket{B}.\nonumber
\end{eqnarray}
Note that the transposition $C^T$ is defined through $\langle l|C^T|k\rangle=\langle k|C|l\rangle$, and depends on the basis.

\item For any two observables $A$ and $B$ acting on $\mathcal{H}$, we have
\begin{equation}
\Ket{A\otimes B} = \Ket{A}\otimes \Ket{B}.
\label{id-tensor-2}
\end{equation}
That is, vectorization of tensor product results in the tensor product of vectorized items.

\textit{Proof.}
Note that $A\otimes B=\sum_{ijkl}A_{ij}|i\rangle\langle j|\otimes B_{kl}|k\rangle\langle l|$. Thus
\begin{eqnarray}
\Ket{A\otimes B} &=&\sum_{ij,kl} A_{ij} |ij\rangle B_{kl}|kl\rangle\nonumber\\
&=& \Ket{A}\otimes \Ket{B}.
\end{eqnarray}
As a result, we have $\Ket{A^{\otimes N}}=\Ket{A}^{\otimes N}$.

\textit{Remark 1.---}For a review of purification methods and their applications in quantum information see Refs. [23].

\end{enumerate}
\section{Proof of Eqs. (5)--(7)}

From $\partial_{\tau}\Ket{\varrho}=\widetilde{\mathcal{L}}_{\tau}(\mathrm{x})\Ket{\varrho}$ and $\widetilde{\mathcal{L}}_{\tau}(\mathrm{x})=\mathrm{x}(\tau)\widetilde{\mathsf{L}}$, we have
\begin{align}
\partial_{\mathrm{x}}\Ket{\varrho}&=\frac{1}{\partial_{\tau}\mathrm{x}(\tau)}\partial_{\tau}\Ket{\varrho}\nonumber\\
&= \frac{1}{\partial_{\tau}\ln \mathrm{x}(\tau)}\widetilde{\mathsf{L}}\Ket{\varrho}.
\end{align}
Hence for the pure state $\widetilde{\varrho}=\KetBra{\varrho}{\varrho}/\mathrm{Tr}[\varrho^2]$ we obtain
\begin{align}
L_{\widetilde{\varrho}}&=2\partial_{\mathrm{x}}\widetilde{\varrho}\label{lwr}\\
&= 2 \Big( \frac{\partial_{\mathrm{x}}\Ket{\varrho}\Bra{\varrho}}{\mathrm{Tr}[\varrho^2]} + \frac{\Ket{\varrho} \partial_{\mathrm{x}}\Bra{\varrho}}{\mathrm{Tr}[\varrho^2]}  -\KetBra{\varrho}{\varrho}\frac{\partial_{\mathrm{x}} \mathrm{Tr}[\varrho^2]}{\mathrm{Tr}[\varrho^2]^2}\Big)\label{int-eq-1}\\
&= 2\Big(\frac{1}{\partial_{\tau}\ln \mathrm{x}(\tau)} \widetilde{\mathsf{L}} \widetilde{\varrho} + \frac{1}{\partial_{\tau}\ln \mathrm{x}(\tau)} \widetilde{\varrho} \widetilde{\mathsf{L}}^{\dag} - \widetilde{\varrho} ~\partial_{\mathrm{x}}\ln \mathrm{Tr}[\varrho^2]\Big). \label{Aa}
\end{align}
As a result, the associated QFI to $\widetilde{\varrho}$ becomes
\begin{align}
\widetilde{\mathcal{F}}^{(\mathrm{Q})} &= \mathrm{Tr}[\widetilde{\varrho}L_{\widetilde{\varrho}}^2]\nonumber\\
&\overset{\text{Eq.~(\ref{Aa})}}{=} \frac{4}{\big[\partial_{\tau}\ln\mathrm{x}(\tau)\big]^2}\mathrm{Cov}_{\widetilde{\varrho}} (\widetilde{\mathsf{L}}^{\dagger},\widetilde{\mathsf{L}}).
\label{SM:FFtilde}
\end{align}
This is Eq.~(5) of the main text.

To relate $\widetilde{\mathcal{F}}^{(\mathrm{Q})}_{\widetilde{\varrho}}$ with $\mathcal{F}^{(\mathrm{Q})}_{\varrho}$, we first need to relate $L_{\widetilde{\varrho}}$ with $L_{\varrho}$. In so doing, we first vectorize the very definition of the SLD $L_{\varrho}$
\begin{equation}
(1/2)(L_{\varrho} \varrho+\varrho L_{\varrho})=\partial_{\mathrm{x}}\varrho \overset{\text{Eq.~(\ref{id-tensor})}}{\Rightarrow} (L_{\varrho}\otimes \openone + \openone \otimes L_{\varrho}^T)\Ket{\varrho}=2\partial_{\mathrm{x}}\Ket{\varrho}.
\label{eq:LL}
\end{equation}
Replacing this into Eq.~(\ref{int-eq-1}) yields
\begin{align}
L_{\widetilde{\varrho}} &= (L_{\varrho}\otimes \openone + \openone \otimes L_{\varrho}^T -\partial_{\mathrm{x}} \ln\mathrm{Tr}[\varrho^2])\widetilde{\varrho} +\widetilde{\varrho} (L_{\varrho}\otimes \openone + \openone \otimes L_{\varrho}^T -\partial_{\mathrm{x}} \ln\mathrm{Tr}[\varrho^2])\nonumber\\
& \overset{\text{Eq.~(\ref{lwr})}}{=} 2\partial_{\mathrm{x}} \widetilde{\varrho}.
\end{align}
Hence by comparing the above relation with the very definition of $L_{\widetilde{\varrho}}$, $\partial_{\mathrm{x}} \widetilde{\varrho} =(1/2)(L_{\widetilde{\varrho}} \widetilde{\varrho} + \widetilde{\varrho} L_{\widetilde{\varrho}})$, we obtain
\begin{align}
L_{\widetilde{\varrho}} &= L_{\varrho}\otimes \openone + \openone \otimes L_{\varrho}^T -\partial_{\mathrm{x}} \ln\mathrm{Tr}[\varrho^2] \label{Ltilde1}\\
&=  L_{\varrho}\otimes \openone + \openone \otimes L_{\varrho}^T - 2\frac{\mathrm{Tr}[\varrho^2 L_{\varrho}]}{\mathrm{Tr}[\varrho^2]},
\label{Ltilde2}
\end{align}
where we have used
\begin{align}
\partial_{\mathrm{x}}\ln\mathrm{Tr}[\varrho^2] =\frac{1}{\mathrm{Tr}[\varrho^2]}\mathrm{Tr}[\partial_{\mathrm{x}}\varrho\varrho+\varrho\partial_{\mathrm{x}}\varrho] \overset{\text{Eq.~(\ref{eq:LL})}}{=} 2\frac{\mathrm{Tr}[\varrho^2 L_{\varrho}]}{\mathrm{Tr}[\varrho^2]}.
\label{beta}
\end{align}

Recalculating $\widetilde{\mathcal{F}}^{(\mathrm{Q})}$ from $L_{\widetilde{\varrho}}$ of Eq.~(\ref{Ltilde2}) yields
\begin{align}
\widetilde{\mathcal{F}}^{(\mathrm{Q})} &=\mathrm{Tr}[\widetilde{\varrho} L_{\widetilde{\varrho}}^2]\label{SM:FF}\\
&=\frac{1}{\mathrm{Tr}[\varrho^2]}\Bra{\varrho}\Big( L_{\varrho}\otimes \openone+\openone\otimes L_{\varrho}^T- 2\mathrm{Tr}[\varrho^2 L_{\varrho}]/\mathrm{Tr}[\varrho^2]\Big)^2\Ket{\varrho}\nonumber\\
&\overset{\text{Eq.~(\ref{id-tensor})}}{=}\frac{1}{\mathrm{Tr}[\varrho^2]}\BraKet{ L_{\varrho}\varrho+\varrho L^T_{\varrho} -2\varrho~\mathrm{Tr}[\varrho^2 L_{\varrho}]/\mathrm{Tr}[\varrho^2]}{ L_{\varrho}\varrho+\varrho L^T_{\varrho} -2\varrho~\mathrm{Tr}[\varrho^2 L_{\varrho}]/\mathrm{Tr}[\varrho^2]}\nonumber\\
&\overset{\text{Eq.~(\ref{id-tr})}}{=} \frac{1}{\mathrm{Tr}[\varrho^2]} \mathrm{Tr}\Big[\big( L_{\varrho}\varrho+\varrho L_{\varrho}- 2\varrho~\mathrm{Tr}[\varrho^2 L_{\varrho}]/\mathrm{Tr}[\varrho^2]\big)^2\Big]\nonumber\\
&= \frac{2}{\mathrm{Tr}[\varrho^2]}\Big( \mathrm{Tr}[\varrho  L_{\varrho}\varrho  L_{\varrho}] + \mathrm{Tr}[\varrho^2  L^2_{\varrho}] -2\frac{(\mathrm{Tr}[\varrho^2  L_{\varrho}])^2}{\mathrm{Tr}[\varrho^2]}\Big),
\label{ftilde-f}
\end{align}
which is Eq.~(6) of the main text.

We now employ the fact that for any pair of positive operators $A$ and $B$, $\lambda_{\min}(A)\mathrm{Tr}[B]\leqslant\mathrm{Tr}[AB]\leqslant \lambda_{\max}(A)\mathrm{Tr}[B]$. Thus for the first term in Eq.~(\ref{ftilde-f}) we obtain
\begin{equation}
\lambda_{\min}(\varrho)\mathcal{F}^{(\mathrm{Q})}\leqslant\mathrm{Tr}[\varrho L_{\varrho}\varrho L_{\varrho}]\leqslant \lambda_{\max}(\varrho)\mathcal{F}^{(\mathrm{Q})},
\label{b-1}
\end{equation}
and similarly (by using $\mathrm{Tr}[\varrho^2 L^2_{\varrho}]=\mathrm{Tr}[\varrho \sqrt{\varrho} L^2_{\varrho}\sqrt{\varrho} ]$) for the second term of Eq.~(\ref{ftilde-f})
\begin{equation}
\lambda_{\min}(\varrho)\mathcal{F}^{(\mathrm{Q})}\leqslant\mathrm{Tr}[\varrho^2 L_{\varrho}^2]\leqslant \lambda_{\max}(\varrho)\mathcal{F}^{(\mathrm{Q})}.
\label{b-2}
\end{equation}
Combining Eqs.~(\ref{ftilde-f}), (\ref{b-1}), and (\ref{b-2}) gives
\begin{equation}
\frac{4\lambda_{\min}(\varrho)}{\mathrm{Tr}[\varrho^2]}\left(\mathcal{F}^{(\mathrm{Q})}-\frac{\mathrm{Tr}[\varrho^2 L_{\varrho}]^2}{\lambda_{\min}(\varrho)\mathrm{Tr}[\varrho^2]}\right)\leqslant \widetilde{\mathcal{F}}^{(\mathrm{Q})} \leqslant \frac{4\lambda_{\max}(\varrho)}{\mathrm{Tr}[\varrho^2]}\left(\mathcal{F}^{(\mathrm{Q})}-\frac{\mathrm{Tr}[\varrho^2 L_{\varrho}]^2}{\lambda_{\max}(\varrho)\mathrm{Tr}[\varrho^2]}\right),
\end{equation}
if $\lambda_{\min}(\varrho)>0$, which in turn can be equivalently rearranged as
\begin{equation}
\frac{\mathrm{Tr}[\varrho^2]}{4\lambda_{\max}(\varrho)} \widetilde{\mathcal{F}}^{(\mathrm{Q})} +\frac{\mathrm{Tr}[\varrho^2 L_{\varrho}]^2}{\lambda_{\max}(\varrho)\mathrm{Tr}[\varrho^2]} \leqslant \mathcal{F}^{(\mathrm{Q})}  \leqslant \frac{\mathrm{Tr}[\varrho^2]}{4\lambda_{\min}(\varrho)} \widetilde{\mathcal{F}}^{(\mathrm{Q})}+\frac{\mathrm{Tr}[\varrho^2 L_{\varrho}]^2}{\lambda_{\min}(\varrho)\mathrm{Tr}[\varrho^2]},
\end{equation}
or simply as
\begin{equation}
\frac{\mathrm{Tr}[\varrho^2]} {4\lambda_{\max}(\varrho)}\widetilde{\mathcal{F}}^{(\mathrm{Q})} \leqslant \mathcal{F}^{(\mathrm{Q})}  \leqslant \frac{\mathrm{Tr}[\varrho^2]}{4\lambda_{\min}(\varrho)} \widetilde{\mathcal{F}}^{(\mathrm{Q})}+\frac{\mathrm{Tr}[\varrho^2 L_{\varrho}]^2}{\lambda_{\min}(\varrho)\mathrm{Tr}[\varrho^2]}.
\label{eq:7}
\end{equation}
This is Eq.~(7) of the main text.

\section{Proof that for a fully product scenario $\widetilde{\mathcal{F}}^{(\mathrm{Q})}(\mathrm{x},\tau;N)=N \widetilde{\mathcal{F}}^{(\mathrm{Q})}(\mathrm{x},\tau;1)$}

For a \textit{separable} or fully product estimation scenario with product input states ($\varrho^{(1)~\otimes N}$) and separate channels, we have $L_{\varrho}=\sum_{m=1}^{N} L^{(m)}_{\varrho^{(1)}}$, where $L^{(m)}_{\varrho^{(1)}}$ is the SLD when the single-probe input state for channel $m$ is $\varrho^{(1)}$. This property yields
\begin{align}
\mathcal{F}^{(\mathrm{Q})}(N) &=\mathrm{Tr}[\varrho L^2_{\varrho}]\nonumber\\
&=\mathrm{Tr}\big[\sum_{m,m'=1}^{N} L_{\varrho_m^{(1)}} L_{\varrho_{m'}^{(1)}}\varrho^{(1)~\otimes N}\big] \nonumber\\
& = \sum_{m=1}^{N} \mathrm{Tr}[\varrho_m^{(1)} L^2_{\varrho_{m}^{(1)}}] \mathrm{Tr}[\varrho^{(1)~\otimes (N-1)}] + \sum_{m\neq m'} \mathrm{Tr}[L_{\varrho_m^{(1)}} \varrho_m^{(1)}] \mathrm{Tr}[L_{\varrho_{m'}^{(1)}} \varrho_{m'}^{(1)}] \mathrm{Tr}[\varrho^{(1)~\otimes (N-2)}\big] \nonumber\\
& = N\mathcal{F}^{(\mathrm{Q})}(1), \label{F:prod}
\end{align}
where in the last line we have used the fact that $\mathrm{Tr}[L_{\xi}\xi]=0$ for any density matrix $\xi$, which is an immediate consequence of the definition of SLD.

Property (\ref{F:prod}) naturally carries over to $\widetilde{F}^{(\mathrm{Q})}$, i.e.,
\begin{equation}
\widetilde{\mathcal{F}}^{(\mathrm{Q})}(\mathrm{x},\tau;N)=N \widetilde{\mathcal{F}}^{(\mathrm{Q})}(\mathrm{x},\tau;1), \label{F=NF}
\end{equation}
if we note that from Eq.~(\ref{id-tensor-2}) a fully product scenario with input state $\varrho^{(1)~\otimes N}$ is equivalent to a fully product scenario with input state $\widetilde{\varrho}_{(1)}^{\otimes N}$. Alternatively, one can also verify validity of Eq.~(\ref{F:prod}) for $\widetilde{\mathcal{F}}^{(\mathrm{Q})}$ through Eq.~(\ref{ftilde-f}).

\section{Example I}

In this example, we have
\begin{align}
\widetilde{\mathsf{L}} &= \sum_{i_1\cdots i_k} \sigma_{i_1}\sigma_{i_2}\cdots \sigma_{i_k}\otimes \sigma_{i_1}\sigma_{i_2}\cdots \sigma_{i_k} -C_{N,k}\openone\otimes\openone,
\end{align}
where we have used $\sigma^{T}=\sigma$ in the computational basis. For odd $k$s we obtain
\begin{align}
\sigma_{i_1}\sigma_{i_2}\cdots \sigma_{i_k}\otimes \sigma_{i_1}\sigma_{i_2}\cdots \sigma_{i_k}(\ket{\Psi}\otimes\ket{\Psi^{\ast}})&=\ket{\Psi^{\perp}}\ket{\Psi^{{\perp}^{\ast}}}, \\
\sigma_{i_1}\sigma_{i_2}\cdots \sigma_{i_k}\otimes \sigma_{i_1}\sigma_{i_2}\cdots \sigma_{i_k}(\ket{\Psi^{\perp}}\otimes\ket{\Psi^{\perp^{\ast}}})&=\ket{\Psi}\ket{\Psi^{\ast}},
\end{align}
where $\ket{\Psi}=(\ket{E_{M}}^{\otimes N}-\ket{E_{m}}^{\otimes N})/\sqrt{2}$ and $\ket{\Psi^{\perp}}=(\ket{E_{M}}^{\otimes N}+\ket{E_{m}}^{\otimes N})/\sqrt{2}$, with $|E_{m}\rangle $ ($|E_{M}\rangle $) being the state corresponding to the smallest (largest) eigenvalue of $\sigma$ (i.e., $\pm1$). This implies that
\begin{equation}
\widetilde{\mathsf{L}}\ket{\Psi}\ket{\Psi^{\ast}}= C_{N,k}(\ket{\Psi^{\perp}}\ket{\Psi^{{\perp}^{\ast}}}-\ket{\Psi}\ket{\Psi^{\ast}}).
\end{equation}
Thus
\begin{align}
\Ket{\varrho(\mathrm{x},\tau)}&=e^{\mathrm{x}\tau\widetilde{\mathsf{L}}}\ket{\Psi}\ket{\Psi^{\ast}}\nonumber\\
&=(1/2)\left([1+e^{-2\mathrm{x}\tau C_{N,k}}]\ket{\Psi}\ket{\Psi^{\ast}} + [-1+ e^{-2\mathrm{x}\tau C_{N,k}}]\ket{\Psi^{\perp}}\ket{\Psi^{\perp^{\ast}}}\right).
\end{align}
Here after we use $\varrho(\tau)$ as a shorthand for $\varrho(\mathrm{x},\tau)$. Now it is straightforward to see
\begin{align}
\mathrm{Tr}[\varrho^2(\tau)]&= (1/2)(1+ e^{-4\mathrm{x}\tau C_{N,k}}),\\
\lambda_{\max}\big(\varrho(\tau)\big)&=(1/2)(1+e^{-2\mathrm{x}\tau C_{N,k}}),\\
\mathcal{K}\big( \varrho(\tau)\big)&= 4\frac{1+ e^{-2\mathrm{x}\tau C_{N,k}}}{1+e^{-4\mathrm{x}\tau C_{N,k}}},\\
\langle\widetilde{\mathsf{L}}\rangle_{\widetilde{\varrho}(\tau)}&= \langle\widetilde{\mathsf{L}}^{\dag}\rangle_{\widetilde{\varrho}(\tau)}=-2C_{N,k}/[1+ e^{-4\mathrm{x}\tau C_{N,k}}],\\
\langle \widetilde{\mathsf{L}}^{\dagger}\widetilde{\mathsf{L}}\rangle_{\widetilde{\varrho}(\tau)}&=4C^2_{N,k}/[1+ e^{-4\mathrm{x}\tau C_{N,k}}],\\
\end{align}
whereby
\begin{equation}
\frac{\widetilde{\mathcal{F}}^{(\mathrm{Q})}(\mathrm{x},\tau;N)}{\mathcal{K}\big( \varrho(\tau)\big)}= \frac{4\tau^2 C_{N,k}^2 e^{-4 \mathrm{x}\tau C_{N,k}}}{[1+\mathrm{e}^{-4\mathrm{x}\tau C_{N,k}}][1+e^{-2\mathrm{x}\tau C_{N,k}}]},
\end{equation}
which is Eq.~(12) of the main text.

\section{Example II}

The master equation
\begin{equation}
\partial_{\tau}\varrho=i\mathrm{x}_1[H,\varrho]+ (1/2)\mathrm{x}_2(\tau)\big(\sum_{m=1}^N\sigma_m^z\varrho\sigma_m^z-N\varrho\big),
\end{equation}
with $H=\sum_{m=1}^{N}\ket{1}_{m}\bra{1}\equiv\sum_m H_m$, is vectorized as
\begin{equation}
\label{differential}
\partial_{\tau}\Ket{\varrho} \overset{\text{Eq.~(\ref{id-tensor})}}{=}\left(i\mathrm{x}_1 \widetilde{\mathsf{L}}_1+ \mathrm{x}_2(\tau) \widetilde{\mathsf{L}}_2\right)\Ket{\varrho},
\end{equation}
where
\begin{align}
&\widetilde{\mathsf{L}}_1=H\otimes \openone-\openone\otimes H^T\equiv \sum_m \widetilde{\mathsf{L}}^{(m)}_1,\\
&\widetilde{\mathsf{L}}_2=(1/2)\big(\sum_{m\in \mathrm{odd}}^{2N-1}\sigma_m^z\otimes\sigma_{m+1}^z-N\openone\big)\equiv \sum_m\widetilde{\mathsf{L}}^{(m)}_2.
\label{eq:L2}
\end{align}
This equation can be solved as follows:
\begin{equation}
\Ket{\varrho(\mathrm{x}_1,\mathrm{x}_2,\tau)} = \mathrm{e}^{i\mathrm{x}_1\tau\widetilde{\mathsf{L}}_1+\Gamma(\tau) \widetilde{\mathsf{L}}_2}\Ket{\varrho(0)},
\label{eq:sol-ex2}
\end{equation}
where $\Gamma(\tau) =\int_0^\tau \mathrm{x}_2(s) \mathrm{d}s$. For brevity, hereafter we replace $\varrho(\mathrm{x}_1,\mathrm{x}_2,\tau)$ with $\varrho(\tau)$.

\subsection{Estimation of $\mathrm{x}_1$}
\subsubsection{Initial product state}

In this case we have a scenario with separate channels with an initial product state given by
\begin{equation}
\Ket{\varrho_{\mathrm{p}}(0)} \overset{\text{Eq.~(\ref{id-tensor-2})}}{=}\Ket{\varrho_{\mathrm{p}}^{(1)}(0)}^{\otimes N},
\label{input-sep}
\end{equation}
where
\begin{align}
\Ket{\varrho_{\mathrm{p}}^{(1)}(0)} &=(1/2)(\ket{00}+\ket{01}+\ket{10}+\ket{11})\nonumber\\
&\overset{.}{=}(1/2)(1~~1~~1~~1)^T,
\end{align}
where the state in the first line is $\in\mathcal{H}^{\otimes 2}_1\otimes \mathcal{H}^{\otimes 2}_2\otimes \ldots \mathcal{H}^{\otimes 2}_N$. The goal is thus to compute $\widetilde{\mathcal{F}}^{(\mathrm{Q})}(\mathrm{x},\tau;1)$ according to Eq.~(5) of the main text. It is straightforward to see that
\begin{align}
\Ket{\varrho_{\mathrm{p}}^{(1)}(\tau)} &\overset{\text{Eq.~(\ref{eq:sol-ex2})}}{=}(1/2)\big(1~~e^{i\mathrm{x}_1\tau-\Gamma(\tau)}~ ~e^{-i\mathrm{x}_1\tau-\Gamma(\tau)}~~1\big)^T,
\label{product-final-state}\\
\varrho_{\mathrm{p}}^{(1)}(\tau)&=(1/2)\left(
\begin{array}{lr}
1& e^{i\mathrm{x}_1\tau-\Gamma(\tau)}\\ e^{-i\mathrm{x}_1\tau-\Gamma(\tau)}&  1\end{array}\right),\\
\lambda_{\max}\big(\varrho^{(1)}_{\mathrm{p}}(\tau)\big)&=e^{-\Gamma(\tau)/2}\mathrm{ch}\big(\Gamma(\tau)/2\big),
\label{lmax-pro-dephasing}\\
\mathrm{Tr}[{\varrho_{\mathrm{p}}^{(1)}}^2(\tau)]&=e^{-\Gamma(\tau)}\mathrm{ch}\big(\Gamma(\tau) \big),\\
 \mathcal{K}\big(\varrho^{(1)}_{\mathrm{p}}(\tau)\big)&=4e^{\Gamma(\tau)/2}\mathrm{ch}\big(\Gamma(\tau)/2\big)/ \mathrm{ch}\big(\Gamma(\tau)\big), \label{K-Product}\\
\widetilde{\mathsf{L}}^{(1)}_1\Ket{\varrho_{\mathrm{p}}^{(1)}(\tau)} &\overset{\cdot}{=}(1/2)\big(0~~-e^{i\mathrm{x}_1\tau-\Gamma(\tau)}~~ e^{-i\mathrm{x}_1\tau-\Gamma(\tau)}~~0\big)^T,\\
\langle\widetilde{\mathsf{L}}^{(1)}_1\rangle_{\widetilde{\varrho}^{(1)}_{\mathrm{p}}(\tau)}&= \langle\widetilde{\mathsf{L}}^{(1)\dag}_1\rangle_{\widetilde{\varrho}^{(1)}_{\mathrm{p}}(\tau)}=0,\\
\langle\widetilde{\mathsf{L}}^{(1)\dag}_1 \widetilde{\mathsf{L}}^{(1)}_1\rangle_{\widetilde{\varrho}^{(1)}_{\mathrm{p}}(\tau)} &=e^{-\Gamma(\tau)}/[2\mathrm{ch}\big(\Gamma(\tau)\big)].
\end{align}
From Eq.~(\ref{SM:FFtilde}) [Eq. (3) of the main text] it can be read
\begin{equation}
\widetilde{\mathcal{F}}^{(\mathrm{Q})}_{\mathrm{p}}(\mathrm{x}_1,\tau;1)=2\tau^2 e^{-\Gamma(\tau)}/\mathrm{ch}\big(\Gamma(\tau)\big),
\end{equation}
and in turn
\begin{align}
\frac{\widetilde{\mathcal{F}}^{(\mathrm{Q})}_{\mathrm{p}}(\mathrm{x_1},\tau;N)}{\mathcal{K}\big(\varrho_{\mathrm{p}}^{(1)}(\tau)\big)} \overset{\text{Eq.~(\ref{F=NF})}}{=}\frac{N\tau^2 e^{-3\Gamma(\tau)/2}}{2\mathrm{ch}\big(\Gamma(\tau)/2\big)},
\label{b-qfi-p-dephasing-delta}
\end{align}
which is Eq.~(13) of the main text.

\begin{figure}[tp]
\includegraphics[scale=.7]{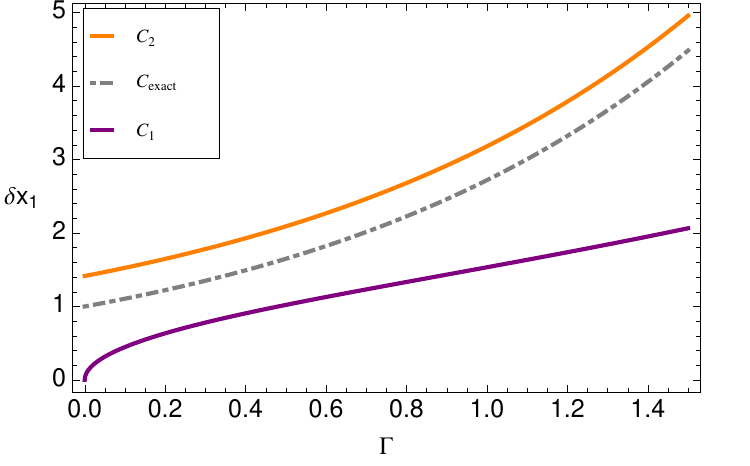}
\caption{(color online.) Constant factor $C(\Gamma)$ in the scaling $O(C/\sqrt{N})$ of the $\mathrm{x}_1$ estimation with product states. The values $C_1$ [down, purple], $C_{\mathrm{exact}}$ [middle, dot-dashed], and $C_2$ [up, orange] are given through Refs. [10,19], exact calculation, and our method, respectively.}
\label{fig:ex-2-comp}
\end{figure}

\textit{Remark 2.}---The very problem of estimating the parameter $\mathrm{x}_1$ has already been discussed in Refs.~[10,19], albeit with different methodologies. Although our method does not compare with the methods in these references, it is interesting to compare the obtained scalings for $\delta \mathrm{x}_1$ with those of Refs.~[10,19]. As argued in Table $1$ of Ref.~[19], the expected scaling for the error should be $O(C/\sqrt{N})$---the very shot-noise scaling---with $C_1=\sqrt{(1-\eta^2)/\eta}$ for both the methods of Ref.~[10] and Ref.~[19]. Here, the parameter $\eta$ can be related to our parameter $\Gamma$ through $\eta=e^{-\Gamma}$. On the other hand, the exact solution gives $C_{\mathrm{exact}}=e^{\Gamma}$ [Eq.~(15) of the main text], whereas our method gives $C_{2}=\sqrt{2e^{3\Gamma/2}\mathrm{ch}(\Gamma/2)}$ [Eq.~(\ref{b-qfi-p-dephasing-delta})]. Figure \ref{fig:ex-2-comp} depicts these results vs. $\Gamma$. It is seen that $C_1$ is always below $C_{\mathrm{exact}}$, while as expected our $C_2$ is above all. Interestingly, $C_1$ does not compare well with $C_{\mathrm{exact}}$ (whether for small or large values of $\Gamma$); however, our $C_{2}$ behaves not only similarly to $C_{\mathrm{exact}}$ for all values of $\Gamma$, it also is closer to this exact value than $C_2$.

\subsubsection{Initial entangled state}

In this case we have $\varrho(0)=|\Psi_{\mathrm{e}}\rangle \langle\Psi_{\mathrm{e}}|$, where $|\Psi_{\mathrm{e}}\rangle=(|0\rangle^{\otimes N}+|1\rangle^{\otimes N})/\sqrt{2}$. Hence
\begin{align}
\Ket{\varrho_{\mathrm{e}}(0)} &=(1/2)\big(\ket{00}^{\otimes N}+\ket{01}^{\otimes N}+\ket{10}^{\otimes N}+\ket{11}^{\otimes N}) \label{input-ent}\\
&\overset{\cdot}{=}(1/2)\big(1~~\bm{0}~~1~~\bm{0}~~1~~\bm{0}~~1\big)^T,
\end{align}
where the state in the first line is $\in\mathcal{H}^{\otimes 2}_1\otimes \mathcal{H}^{\otimes 2}_2\otimes \ldots \mathcal{H}^{\otimes 2}_N$. For the ease of notation hereafter we work in the subspace excluding those marked with $\bm{0}$s above. It is then straightforward to see that
\begin{align}
\Ket{\varrho_{\mathrm{e}}(\tau)}& \overset{\text{Eq.~(\ref{eq:sol-ex2})}}{=} (1/2)\big(
1~~ e^{iN\mathrm{x}_1 \tau-N\Gamma(\tau)}~~e^{-iN\mathrm{x}_1 \tau-N\Gamma(\tau)}~~1\big)^T\label{rho-e}\\
\varrho_{\mathrm{e}}(\tau)&= (1/2)\left(
\begin{array}{lr}
 1 & e^{iN\mathrm{x}_1 \tau-N\Gamma(\tau)} \\
 e^{-iN\mathrm{x}_1 \tau-N\Gamma(\tau)} & 1\end{array}\right),\\
\lambda_{\max}\big(\varrho_{\mathrm{e}}(\tau)\big) &= e^{-N\Gamma(\tau)/2}\mathrm{ch}\big(N \Gamma(\tau)/2\big), \label{lmax-e-D} \\
\mathrm{Tr}[\varrho^2_{\mathrm{e}}(\tau)]& =e^{-N\Gamma(\tau)}\mathrm{ch}\big(N\Gamma(\tau) \big), \label{trace2-e-D}\\
\mathcal{K}\big( \varrho_{\mathrm{e}}(\tau)\big)&=4e^{N\Gamma(\tau)/2}\mathrm{ch}\big(N\Gamma(\tau)/2\big)/\mathrm{ch}\big(N\Gamma(\tau)\big), \label{K-Entangled}\\
\widetilde{\mathsf{L}}_1\Ket{\varrho_\mathrm{e}(\tau)}&=(1/2)\big[\sum_{m\in\mathrm{odd}}^{2N} \ket{1}_m\bra{1}\otimes \openone- \sum_{m\in\mathrm{even}}^{2N}\openone\otimes \ket{1}_m\bra{1}\big]\times\nonumber\\
&~~~~~~ (\ket{00}^{\otimes N}+e^{iN\mathrm{x}_1 \tau-N\Gamma(\tau)} \ket{01}^{\otimes N}+ e^{-iN \mathrm{x}_1 \tau-N\Gamma(\tau)}\ket{10}^{\otimes N} + \ket{11}^{\otimes N})\nonumber\\
&\overset{\cdot}{=}(N/2)\big(0~~-e^{iN\mathrm{x}_1 \tau-N\Gamma(\tau)} ~~e^{-iN\mathrm{x}_1 \tau-N\Gamma(\tau)}~~0\big)^T,\\
\langle \widetilde{\mathsf{L}}_1\rangle_{\widetilde{\varrho}_{\mathrm{e}}(\tau)}&= \langle \widetilde{\mathsf{L}}^{\dag}_1\rangle_{\widetilde{\varrho}_{\mathrm{e}}(\tau)}= 0,\\
\langle\widetilde{\mathsf{L}}_1^{\dagger}~\widetilde{\mathsf{L}}_1\rangle_{\widetilde{\varrho}_{\mathrm{e}}(\tau)} &=N^2 e^{-2N\Gamma(\tau)}/[2~\mathrm{ch}\big( N\Gamma(\tau)\big)].
\end{align}
Thus
\begin{equation}
\frac{\widetilde{\mathcal{F}}^{(\mathrm{Q})}_{\mathrm{e}}(\mathrm{x}_1,\tau;N)}{\mathcal{K}\big( \varrho_{\mathrm{e}}(\tau)\big)}=\frac{N^2\tau^2 e^{-3N\Gamma(\tau)/2}}{2~\mathrm{ch}\big(N\Gamma(\tau)/2\big)},
\end{equation}
which is Eq.~(14) of the main text.

\textit{Remark 3.---}One can provide a further comparison of the performance of the separable scenario (using product states) with the entangled scenario (using entangled states) by taking into account the best \textit{interrogation} times for both scenarios (see Refs. [22,29] of the main text for analogous comparisons). Such comparison may prove useful for practical purposes.

Let us assume that the total experiment (measurement) time for each scenario is $T$, during which we perform measurements in each interrogation time $\tau$. Thus $M=T/\tau$ shows how many times we have repeated the scenarios. For the separable scenario, we choose $N$ independent systems as in Eq.~(\ref{input-sep}) as our input; while for the entangled scenario, we choose the $N$-partite entangled state (\ref{input-ent}) as the input. Now we define the relative resolution $R$ as the ratio of the \textit{optimized} bounds on the accuracies of the separable and entangled scenarios over the interrogation times ($\tau_{\mathrm{p}}$ and $\tau_{\mathrm{e}}$),
\begin{align}
R^2 & \equiv \frac{\min_{\tau}{\left(\delta \mathrm{x}_1\right)^2}|_{\mathrm{p}}}{\min_{\tau}{\left(\delta \mathrm{x}_1\right)^2}|_{\mathrm{e}}}=\frac{ \max_{\tau} M_{\mathrm{e}}\widetilde{\mathcal{F}}^{(\mathrm{Q})}_{\mathrm{e}}(\mathrm{x}_1,\tau;N)/\mathcal{K}\big( \varrho_{\mathrm{e}}(\tau)\big)}{ \max_{\tau} M_{\mathrm{p}}\widetilde{\mathcal{F}}^{(\mathrm{Q})}_{\mathrm{p}}(\mathrm{x}_1,\tau;N)/\mathcal{K}\big( \varrho^{(1)}_{\mathrm{p}}(\tau)\big)}\\
&=\frac{1}{N} \frac{ \max_{\tau} \widetilde{\mathcal{F}}^{(\mathrm{Q})}_{\mathrm{e}}(\mathrm{x}_1,\tau;N)/\big[\tau\mathcal{K}\big( \varrho_{\mathrm{e}}(\tau)\big)\big]}{ \max_{\tau} \widetilde{\mathcal{F}}^{(\mathrm{Q})}_{\mathrm{p}}(\mathrm{x}_1,\tau;1)/\big[\tau\mathcal{K}\big( \varrho^{(1)}_{\mathrm{p}}(\tau)\big)\big]}\\
& = N\frac{\tau_{\mathrm{e}}}{\tau_{\mathrm{p}}}e^{-3N\Gamma(\tau_{\mathrm{e}})/2+3\Gamma(\tau_{\mathrm{p}})/2}\frac{\mathrm{ch}\big( \Gamma(\tau_{\mathrm{p}}/2)\big)}{\mathrm{ch}\big( \Gamma(N\tau_{\mathrm{e}}/2)\big)}. \label{R-eq}
\end{align}
The best/optimized interrogation times $\tau_{\mathrm{p}}$ and $\tau_{\mathrm{e}}$ are obtained by the following equations:
\begin{align}
\partial_{\tau}\left(\frac{\widetilde{\mathcal{F}}^{(\mathrm{Q})}_{\mathrm{p}}(\mathrm{x}_1,\tau;1)}{\tau\mathcal{K}\big( \varrho^{(1)}_{\mathrm{p}}(\tau)\big)} \right)\Big|_{\tau=\tau_{\mathrm{p}}} &=0,\\
\partial_{\tau}\left(\frac{\widetilde{\mathcal{F}}^{(\mathrm{Q})}_{\mathrm{e}}(\mathrm{x}_1,\tau;N)}{\tau\mathcal{K}\big( \varrho_{\mathrm{e}}(\tau)\big)} \right)\Big|_{\tau=\tau_{\mathrm{e}}} & =0,
\end{align}
or from
\begin{align}
&\tau_{\mathrm{p}} \big(3+\mathrm{th}[\Gamma(\tau_{\mathrm{p}})/2] \big)\mathrm{x}_2(\tau_{\mathrm{p}})=2, \label{taup}\\
&N\tau_{\mathrm{e}}  \big(3+\mathrm{th}[N\Gamma(\tau_{\mathrm{e}})/2]\big) \mathrm{x}_2(\tau_{\mathrm{e}})=2, \label{taue}
\end{align}
where $\mathrm{th}\equiv \tanh$.

In the absence of dephasing ($\Gamma=0$), we have $\tau_{\mathrm{p}}= \tau_{\mathrm{e}}$, from whence $R^2=N$. This is the very Heisenberg limit for closed systems. In the Markovian case, we have $\mathrm{x}_2(\tau)=\mathrm{x}_2 \tau$ (i.e., $\Gamma(\tau)=\mathrm{x}_2 \tau$). Replacing the solutions of Eqs.~(\ref{taup}) and (\ref{taue}) [$\tau_{\mathrm{p}} \mathrm{x}_2 = N\tau_{\mathrm{e}} \mathrm{x}_2\approx0.607065$] in Eq.~(\ref{R-eq}) yields that here $R^2=1$. That is, in the Markovian case the entangled scenario for the best corresponding interrogation time does not offer an advantage over the separable scenario in its best interrogation time. This is an important result that perhaps defies the common expectation that the entangled scenario should in general outperform the separable scenario. Nonetheless, in the non-Markovian case the entangled scenario may be relatively advantageous over the separable scenario in some dynamical regimes. An exhaustive precursor investigation of this sort can be found in Ref. [22].

\subsection{Estimation of $\mathrm{x}_2$}

Here the only thing that is different from the case of the estimation of $\mathrm{x}_1$ is the generator $\widetilde{\mathsf{L}}_2$ [Eq.~(\ref{eq:L2})]. The other parameters are all the same as in the estimation of $\mathrm{x}_1$, for both product and entangled initial states.

\subsubsection{Initial product state}

From Eq.~(\ref{product-final-state}) we obtain
\begin{align}
\widetilde{\mathsf{L}}^{(1)}_2\Ket{\varrho_{\mathrm{p}}^{(1)}(\tau)}&\overset{\cdot}{=} -(1/2)\big(
0~~ e^{i\mathrm{x}_1\tau-\Gamma(\tau)}~~ e^{-i\mathrm{x}_1\tau-\Gamma(\tau)}~~0\big),\\
\langle \widetilde{\mathsf{L}}^{(1)}_2\rangle_{\widetilde{\varrho}_{\mathrm{p}}^{(1)}(\tau)} &= - e^{-\Gamma(\tau)}/[2~\mathrm{ch}\big( \Gamma(\tau)\big)],\\
\langle \widetilde{\mathsf{L}}_2^{(1)\dagger}~\widetilde{\mathsf{L}}^{(1)}_2\rangle_{\widetilde{\varrho}_{\mathrm{p}}^{(1)}(\tau)} &=e^{-\Gamma(\tau)}/[2~\mathrm{ch}\big( \Gamma(\tau)\big)],
\end{align}
whence
\begin{align}
\widetilde{\mathcal{F}}_\mathrm{p}^{(\mathrm{Q})}(\mathrm{x}_2,\tau;N) \overset{}{=} N\tau^2/\mathrm{ch}^2\big( \Gamma(t)\big),
\end{align}
and
\begin{align}
\frac{\widetilde{\mathcal{F}}_{\mathrm{p}}^{(\mathrm{Q})}(\mathrm{x}_2,\tau;N)}{\mathcal{K}\big( \varrho^{(1)}_{\mathrm{p}}(\tau)\big)} =
\frac{N e^{-\Gamma(\tau)/2}}{4[\partial \tau\ln \mathrm{x}_2(\tau)]^2 ~\mathrm{ch}\big(\Gamma(\tau)\big) \mathrm{ch}\big(\Gamma(\tau)/2\big)},
\label{b-qfi-p-G}
\end{align}
which is Eq.~(17) of the main text.

\subsubsection{Initial entangled state}

Here from Eqs.~(\ref{rho-e}), (\ref{trace2-e-D}), and (\ref{K-Entangled}) we obtain
\begin{align}
\widetilde{\mathsf{L}}_2\Ket{\varrho_{\mathrm{e}}(\tau)} &=(1/4)\big(\sum_{m\in \mathrm{odd}}^{2N-1}\sigma_m^z\otimes\sigma_{m+1}^z-N\openone\big)	\times\nonumber\\
&\big(\Ket{00}^{\otimes N}+ e^{iN\mathrm{x}_1 \tau-N\Gamma(\tau)}\Ket{01}^{\otimes N} + e^{-iN \mathrm{x}_1 \tau-N\Gamma(\tau)} \Ket{10}^{\otimes N}+ \Ket{11}^{\otimes N}\big) \nonumber\\
& \overset{\cdot}{=} -(N/2)\big(0~~ e^{iN\mathrm{x}_1 \tau-N\Gamma(\tau)}~~ e^{-iN\mathrm{x}_1 \tau-N\Gamma(\tau)}~~0
\big)^T,\\
\langle\widetilde{\mathsf{L}}_2\rangle_{\widetilde{\varrho}_{\mathrm{e}}(\tau)} &= \langle\widetilde{\mathsf{L}}^{\dag}_2\rangle_{\widetilde{\varrho}_{\mathrm{e}}(\tau)} =-N e^{-2N\Gamma(\tau)}/[2~\mathrm{ch}\big( \Gamma(\tau)\big)],\\
\langle \widetilde{\mathsf{L}}^{\dagger}_2~\widetilde{\mathsf{L}}_2\rangle_{\widetilde{\varrho}_{\mathrm{e}}(\tau)} &= N^2 e^{-N\Gamma(\tau)}/[2~\mathrm{ch}\big( \Gamma(\tau)\big)],
\end{align}
whence
\begin{equation}
\frac{\widetilde{\mathcal{F}}_{\mathrm{e}}^{(\mathrm{Q})}(\mathrm{x}_2,\tau;N)}{\mathcal{K}\big( \varrho_{\mathrm{e}}(\tau)\big)} =\frac{N^2 e^{-N\Gamma(\tau)/2}}{4[\partial \tau\ln \mathrm{x}_2(\tau)]^2\mathrm{ch}\big(N\Gamma(\tau)\big) \mathrm{ch}\big(N\Gamma(\tau)/2\big)},
\end{equation}
which is Eq.~(18) of the main text.

\section{Example III}

Consider a lossy bosonic channel described by
\begin{equation}
\mathcal{L}_{\tau}[\varrho]= \mathrm{x}[a\varrho a^{\dagger} -( \widehat{n}\varrho + \varrho \widehat{n})/2],
\end{equation}
 where $a$ ($a^{\dag}$) is the bosonic annihilation (creation) operator, $\widehat{n}=a^{\dag}a$, and $\mathrm{x}$ is the loss parameter. The Cram\'er-Rao bound for estimation of $\varphi$, defined through $\tan^2[\varphi(\mathrm{x},\tau)] = e^{\mathrm{x} \tau} - 1$, has been obtained as $\delta\varphi\geqslant 1/\sqrt{4\overline{n}\tau}$, and whereby $\delta\mathrm{x}\geqslant \sqrt{\mathrm{x}/(\overline{n}\tau)}$, where $\overline{n}=\mathrm{Tr}[\widehat{n}\varrho(0)]$ [31]. Particularly, it has been shown that Fock states are optimal for this estimation [20]. Here we revisit this example and demonstrate that the scaling of the error is captured correctly in our framework.

The evolution of this system, when the initial state is $\varrho(0)=|N\rangle \langle N|$ (whence $\overline{n}=N$), is given by [20]
\begin{equation}
\varrho(\mathrm{x},\tau)= \sum_{m=0}^{\infty} \frac{1}{m!}\mathrm{s}^{2m}\mathrm{c}^{\widehat{n}}~a^{m}\varrho(0)(a^{\dagger})^m \mathrm{c}^{\widehat{n}},
\end{equation}
or alternatively in the photon number basis
\begin{align}
\varrho(\mathrm{x},\tau)&=\sum_{\ell,\ell'=0}^{\infty}\langle \ell|\varrho(\mathrm{x},\tau)|\ell'\rangle |\ell\rangle \langle \ell'|\nonumber\\
&= \sum_{\ell,\ell'=0}^{\infty} \sum_{m=0}^{\infty} \frac{1}{m!}\mathrm{s}^{2m}\langle \ell|\mathrm{c}^{\widehat{n}}~a^{m}|N\rangle \langle N|(a^{\dagger})^m \mathrm{c}^{\widehat{n}}|\ell'\rangle |\ell\rangle \langle \ell'|\nonumber\\
&= \sum_{\ell,\ell'=0}^{\infty} \sum_{m=0}^{\infty} \frac{1}{m!}\mathrm{s}^{2m} \mathrm{c}^{\ell+\ell'} \sqrt{\frac{(\ell+m)!(\ell'+m)!}{\ell!\ell'!}}\delta_{\ell+m,N}\delta_{\ell'+m,N} |\ell\rangle \langle \ell'|\nonumber\\
&= \sum_{m=0}^{N} \mathrm{s}^{2m} \mathrm{c}^{2(N-m)} C_{N,m}|N-m\rangle \langle N-m|,
\label{dm}
\end{align}
where $\mathrm{s}=\sin\varphi$, $\mathrm{c}=\cos\varphi$, and $C_{N,m}=\binom{N}{m}$. Note that this density matrix is diagonal, from whence
\begin{equation}
\lambda_{\max}\big(\varrho(\mathrm{x},\tau)\big)=\underset{m\in\{0,1,\ldots,N\}}{\max}\mathrm{s}^{2m}\mathrm{c}^{2(N-m)}C_{N,m}.
\end{equation}


Vectorizing Eq.~(\ref{dm}) in the photon number basis gives \footnote{Since Eq.~(\ref{dm}) is the spectral decomposition of $\varrho$, property (\ref{id-spectral}) indicates that the form (\ref{dm:vec}) is indeed the same as the vectorization of $\varrho$ in the very computational basis.}
\begin{equation}
\Ket{\varrho(\mathrm{x}, \tau)} =\sum_{m=0}^{N} \mathrm{s}^{2m}\mathrm{c}^{2(N-m)}C_{N,m}|N-m\rangle |N-m\rangle,
\label{dm:vec}
\end{equation}
and
\begin{equation}
\BraKet{\varrho(\mathrm{x},\tau)}{\varrho(\mathrm{x},\tau)}=\mathrm{Tr}[\varrho^2(\mathrm{x},\tau)]= \sum_{m=0}^{N} \mathrm{s}^{4m} \mathrm{c}^{4(N-m)} C_{N,m}^2.
\end{equation}
We then can construct the (normalized) density matrix $\widetilde{\varrho}(\mathrm{x},\tau)\equiv\Ket{\varrho(\mathrm{x},\tau)}\Bra{\varrho(\mathrm{x},\tau)}/\mathrm{Tr[\varrho^2(\mathrm{x},\tau)]}$.

Similarly, vectorizing $\mathcal{L}_{\tau}$ in the photon number basis yields
\begin{equation}
\widetilde{\mathsf{L}}_3 = a\otimes a-(1/2)(\widehat{n}\otimes \openone +\openone\otimes \widehat{n}),
\end{equation}
where we have used Eq.~(\ref{id-tensor}) with $a^{*}=a$ and $\widehat{n}^T=\widehat{n}^*=\widehat{n}$ (in the photon number basis). Straightforward calculations yield
\begin{align}
\Bra{\varrho(\mathrm{x},\tau)}\widetilde{\mathsf{L}}_3^{\dagger}\Ket{\varrho(\mathrm{x},\tau)}&=\Bra{\varrho(\mathrm{x},\tau)}\widetilde{\mathsf{L}}_3\Ket{\varrho(\mathrm{x},\tau)}=\sum _{m=0}^N \mathrm{s}^{4m}\mathrm{c}^{4(N-m)} C_{N,m}^2 A_{N,m},\label{eq:1}\\
\Bra{\varrho(\mathrm{x},\tau)}\widetilde{\mathsf{L}}_3^{\dagger}\widetilde{\mathsf{L}}_3\Ket{\varrho(\mathrm{x},\tau)}&=\sum _{m=0}^N \mathrm{s}^{4m}\mathrm{c}^{4(N-m)}C_{N,m}^2 A_{N,m}^2,\label{eq:2}
\end{align}
where $A_{N,m}=m(1+\cot^2\varphi)-N$. Now putting everything together, we can calculate
\begin{align}
\mathrm{Cov}_{\widetilde{\varrho}} (\widetilde{\mathsf{L}}_3^{\dagger},\widetilde{\mathsf{L}}_3) & \equiv \mathrm{Tr}[\widetilde{\mathsf{L}}_3^{\dag}\widetilde{\mathsf{L}}_3\widetilde{\varrho}(\mathrm{x},\tau)]- \mathrm{Tr}[\widetilde{\mathsf{L}}_3^{\dag}\widetilde{\varrho}(\mathrm{x},\tau)]\mathrm{Tr}[\widetilde{\mathsf{L}}_3\widetilde{\varrho}(\mathrm{x},\tau)]\nonumber\\
&\overset{\text{(\ref{eq:1}) \& (\ref{eq:2})}}{=}\frac{1}{ \sum_{m=0}^{N} \mathrm{s}^{4m} \mathrm{c}^{4(N-m)} C_{N,m}^2}\left(\sum _{m=0}^N \mathrm{s}^{4m}\mathrm{c}^{4(N-m)} C_{N,m}^2 A_{N,m}^2-\frac{\left(\sum _{m=0}^N \mathrm{s}^{4m}\mathrm{c}^{4(N-m)} C_{N,m}^2 A_{N,m}\right)^2}{ \sum_{m=0}^{N} \mathrm{s}^{4m} \mathrm{c}^{4(N-m)} C_{N,m}^2}\right).
\end{align}
This relation in turn enables us to compute the generalized QFI $\widetilde{\mathcal{F}}^{(\mathrm{Q})}(\mathrm{x},\tau;N)$ defined as
\begin{equation}
\widetilde{\mathcal{F}}^{(\mathrm{Q})}(\mathrm{x},\tau;N)\equiv 4\tau^2 \mathrm{Cov}_{\widetilde{\varrho}} (\widetilde{\mathsf{L}}_2^{\dagger},\widetilde{\mathsf{L}}_2),
\end{equation}
which is Eq. (5) of the main text of the paper. To estimate $\varphi$ we shall need $\widetilde{\mathcal{F}}^{(\mathrm{Q})}(\varphi,\tau;N)$ rather than $\widetilde{\mathcal{F}}^{(\mathrm{Q})}(\mathrm{x},\tau;N)$, which is given by the conversion rule
\begin{equation}
\widetilde{\mathcal{F}}^{(\mathrm{Q})}(\varphi,\tau;N) = \big(\frac{\partial\varphi}{\partial\mathrm{x}}\big)^{-2} \widetilde{\mathcal{F}}^{(\mathrm{Q})}(\mathrm{x},\tau;N),
\end{equation}
where $(\partial\varphi/\partial\mathrm{x})^2=\tau^2/(4\tan^2\varphi)$. The analytic expression for the bound on $(\delta\varphi)_{\min}$ is thus found according to Eq.~(7) of the main text of the paper, as follows:
\begin{align}
(\delta\varphi)^2_{\min}&\equiv \frac{1}{\mathcal{F}^{(\mathrm{Q})} (\varphi,\tau;N)} \nonumber\\
&\leqslant\frac{4\lambda_{\max}\big(\varrho(\mathrm{x},\tau)\big)}{\mathrm{Tr}[\varrho^2 (\mathrm{x},\tau)]\widetilde{\mathcal{F}}^{(\mathrm{Q})} (\varphi,\tau;N)} \nonumber\\
&= \frac{1}{4}\cot^{2}\varphi ~\underset{0\leqslant m\leqslant N}{\max} \big[C_{N,m}\mathrm{s}^{2m} \mathrm{c}^{2(N-m)}\big]\Big[\sum _{m=0}^N \mathrm{s}^{4m}\mathrm{c}^{4(N-m)}C_{N,m}^2 A_{N,m}^2-\frac{\left(\sum _{m=0}^N \mathrm{s}^{4m}\mathrm{c}^{4(N-m)}C_{N,m}^2 A_{N,m}\right)^2}{\sum_{m=0}^{N} \mathrm{s}^{4m} \mathrm{c}^{4(N-m)} C_{N,m}^2}\Big]^{-1}.
\label{eq:formula}
\end{align}
 Figure~2 of the main text depicts this relation for different values of $\varphi$ and $N$, which captures correctly the expected $1/\sqrt{N}$ result (with the same constant factor $1/2$) obtained from the exact calculations for the QFI [20,31].
\end{widetext}


\begin{thebibliography}{90}

\bibitem{Estimation:book} P. R. Bevington and D. K. Robinson, \emph{Data Reduction and Error Analysis for the Physical Sciences} (McGraw-Hill, New York, 2003).

\bibitem{Cramer:book} H. Cram\'{e}r, \emph{Mathematical Methods of Statistics} (Princeton University Press, Princeton, NJ, 1946).

\bibitem{Helstrom:book} C. W. Helstrom, \emph{Quantum Detection and Estimation Theory} (Academic Press, New York, 1976); A. S. Holevo, \emph{Probabilistic and Statistical Aspects of Quantum Theory} (North-Holland, Amsterdam, 1982).

\bibitem{Lloyd-qmetrology:PRL} P. Cappellaro, J. Emerson, N. Boulant, C. Ramanathan, S. Lloyd, and D. G. Cory, Phys. Rev. Lett. \textbf{94}, 020502 (2005);
V. Giovannetti, S. Lloyd, and L. Maccone,
\textit{ibid.}, \textbf{96}, 010401 (2006).

\bibitem{Boixo-etal:PRL07} S. Boixo, S. T. Flammia, C. M. Caves, and JM Geremia, Phys. Rev. Lett. \textbf{98}, 090401 (2007);
M. Napolitano, M. Koschorreck, B. Dubost, N. Behbood, R. J. Sewell, and M. W. Mitchell, Nature \textbf{471}, 486 (2011).

\bibitem{Roy-Braunstein:PRL08} S. M. Roy and S. L. Braunstein, Phys. Rev. Lett. \textbf{100}, 220501 (2008).

\bibitem{Rivas-Luis:PRL10} A. Rivas and A. Luis, Phys. Rev. Lett. \textbf{105}, 010403 (2010).

\bibitem{Kok:PRL10} M. Zwierz, C. A. P\'{e}rez-Delgado, and P. Kok, Phys. Rev. Lett. \textbf{105}, 180402 (2010).

\bibitem{experiments} D. Leibfried, M. D. Barrett, T. Schaetz, J. Britton, J. Chiaverini, W. M. Itano, J. D. Jost, C. Langer, and D. J. Wineland, Science \textbf{304}, 1476 (2004);
G. Brida, M. Genovese, and I. Ruo Berchera, Nature Photon. \textbf{4}, 227 (2010);
B. L\"{u}cke, M. Scherer, J. Kruse, L. Pezz\'{e}, F. Deuretzbacher, P. Hyllus, O. Topic, J. Peise, W. Ertmer, J. Arlt, L. Santos, A. Smerzi, and C. Klempt, Science \textbf{334}, 773 (2011);
M. A. Taylor, J. Janousek, V. Daria, J. Knittel, B. Hage, H.-A. Bachor, and W. P. Bowen, Nature Photon. \textbf{7}, 229 (2013).

\bibitem{Escher:NatPhys} B. M. Escher, R. L. de Matos Filho, and L. Davidovich, Nature Phys. \textbf{7}, 406 (2011).

\bibitem{Watanabe:PRL10} Y. Watanabe, T. Sagawa, and M. Ueda, Phys. Rev. Lett. \textbf{104}, 020401 (2010).

\bibitem{Alicki-Lendi:book} R. Alicki and K. Lendi, \emph{Quantum Dynamical Semigroups and Application} (Springer-Verlag, Berlin, Heidelberg, 1987).

\bibitem{Rivas-Huelga:book} H.-P. Breuer and F. Petruccione, \emph{The Theory of Open Quantum Systems} (Oxford University Press, New York, 2002); A. Rivas and S. F. Huelga, \emph{Open Quantum Systems: An Introduction} (Springer, Heidelberg, 2012).

\bibitem{Lu-Wang-Sun:PRA10} X.-M. Lu, X. Wang, and C. P. Sun, Phys. Rev. A \textbf{82}, 042103 (2010).

\bibitem{Alipour:PRA12} H.-P. Breuer, E.-M. Laine, and J. Piilo, Phys. Rev. Lett. \textbf{103}, 210401 (2009);
A. Rivas, S. F. Huelga, and M. B. Plenio, Phys. Rev. Lett. \textbf{105}, 050403 (2010);
S. Alipour, A. Mani, and A. T. Rezakhani, Phys. Rev. A \textbf{85}, 052108 (2012);
K. Modi, A. Brodutch, H. Cable, T. Paterek, and V. Vedral, Rev. Mod. Phys. \textbf{84}, 1655 (2012).

\bibitem{Fleming-Hu} C. H. Fleming and B. H. Hu, Ann. Phys. \textbf{327}, 1238 (2012).

\bibitem{Pernice:JPB} A. Pernice, J. Helm, and W. T. Strunz, J. Phys. B: At. Mol. Opt. Phys. \textbf{45}, 154005 (2012).

\bibitem{Lloyd:NP} V. Giovannetti, S. Lloyd, and L. Maccone, Nature Photon. \textbf{5}, 222 (2011).

\bibitem{Guta:NatureC} R. Demkowicz-Dobrza\'{n}ski, J. Ko{\l}odynski, and M. Gu\c{t}\u{a}, Nature Commun. \textbf{3}, 1063 (2012).

\bibitem{Adesso:PRA} G. Adesso, F. Dell'Anno, S. De Siena, F. Illuminati, and L. A. M. Souza, Phys. Rev. A \textbf{79},  040305(R) (2009).

\bibitem{Monras:PRA} A. Monras and F. Illuminati, Phys. Rev. A \textbf{83}, 012315 (2011).

\bibitem{Chin:PRL2012} A. W. Chin, S. F. Huelga, and M. B. Plenio, Phys. Rev. Lett. \textbf{109}, 233601 (2012).

\bibitem{Caves-Andersson} C. M. Caves, J. Superconductivity \textbf{12}, 707 (1999);
E. Andersson, J. D. Cresser, and M. J. W. Hall, J. Mod. Opt. \textbf{54}, 1695 (2007).

\bibitem{vec} Vectorization of an operator $A=\sum_{ii'}\langle u_i|A|u_{i'}\rangle |u_{i}\rangle \langle u_{i'}|$ (represented in an orthonormal basis $\{|u_i\rangle\}$) is defined through $\Ket{A}=\sum_{ii'} \langle u_i|A|u_{i'}\rangle |u_{i}\rangle |u^*_{i'}\rangle$, where $|u^*_i\rangle$ is the complex conjugate of $|u_i\rangle$ represented in the computational basis. Two related identities we use most here are $\BraKet{A}{B}=\mathrm{Tr}[A^{\dag}B]$ and $\Ket{AXB}=(A\otimes B^{T})\Ket{X}$, where $T$ denotes transposition in the computational basis. For more details, see Ref.~\cite{SM}.

\bibitem{SM} See Supplemental Materials.

\bibitem{Braunstein-Caves:QFI} S. L. Braunstein and	C. M. Caves, Phys. Rev. Lett. \textbf{72}, 3439 (1994);
S. L. Braunstein, C. M. Caves, and G. J. Milburn, Ann. Phys. (N.Y.) \textbf{247}, 135 (1996);
M. G. A. Paris, Intl. J. Quantum Inf. \textbf{7} (Suppl.), 125 (2009);
D. Braun, Eur. Phys. J. D \textbf{59}, 521 (2010).

\bibitem{Hayashi:book} M. Hayashi, \emph{Quantum Information: An Introduction} (Springer-Verlag, Berlin, Heidelberg, 2006).

\bibitem{Braun:PRL02} D. Braun, Phys. Rev. Lett. \textbf{89}, 277901 (2002).

\bibitem{Benatti:PRL03} F. Benatti, R. Floreanini, and M. Piani, Phys. Rev. Lett. \textbf{91}, 070402 (2003).

\bibitem{Huelga:PRL1997} S. F. Huelga, C. Macchiavello, T. Pellizzari, A. K. Ekert, M. B. Plenio, and J. I. Cirac, Phys. Rev. Lett. \textbf{79},  (1997) 3865.

\bibitem{PRL:chaves} R. Chaves, J. B. Brask, M. Markiewicz, J. Ko{\l}odynski, and A. Ac\'{i}n, Phys. Rev. Lett. \textbf{111}, 120401 (2013).

\bibitem{Campo:PRL2013} A. del Campo, I. L. Egusquiza, M. B. Plenio, and S. F. Huelga, Phys. Rev. Lett. \textbf{110}, 050403 (2013).

\bibitem{Paris:PRL} A. Monras and M. G. A. Paris, Phys. Rev. Lett. \textbf{98}, 160401 (2007).

\bibitem{GLM-new:PRL} V. Giovannetti, S. Lloyd, and L. Maccone, Phys. Rev. Lett. \textbf{108}, 260405 (2012).

\bibitem{opt} J. Abadie \textit{et al.} (The LIGO Scientific Collaboration), Nature Phys. \textbf{7}, 962 (2011).

\end{thebibliography}
\end{document}